\documentstyle[aps]{revtex}

\begin{document}

\preprint{EFUAZ FT-96-31}

\title{Questions in the Theory of \linebreak
the $(1,0)\oplus (0,1)$ Quantized Fields\thanks{Submitted to
``Foundation of Physics"}}

\smallskip

\author{{\bf Valeri V. Dvoeglazov}}

\address {
Escuela de F\'{\i}sica, Universidad Aut\'onoma de Zacatecas \\
Antonio Doval\'{\i} Jaime\, s/n, Zacatecas 98068, Zac., M\'exico\\
Internet address:  VALERI@CANTERA.REDUAZ.MX
}

\date{}

\maketitle

\medskip

\begin{abstract}
\baselineskip13pt

\medskip

We find a mapping between
antisymmetric tensor matter fields and the Weinberg's
$2(2j+1)$- component ``bispinor" fields.
Equations which describe the $j=1$ antisymmetric tensor field
coincide with the Hammer-Tucker equations entirely and with
the Weinberg ones within a subsidiary condition,
the Klein-Gordon equation. The new Lagrangian for the Weinberg theory is
proposed which is scalar and Hermitian. It is built on the basis of
the concept of the `Weinberg doubles'. Origins of  a contradiction between
the classical theory, the Weinberg theorem $B-A=\lambda$  for quantum
relativistic fields and the claimed `longitudity' of the antisymmetric
tensor field (transformed on the $(1,0)\oplus (0,1)$ Lorentz group
representation) after quantization are clarified.  Analogs of the $j=1/2$
Feynman-Dyson propagator are presented in the framework of the $j=1$
Weinberg theory.  It is then shown that under the definite choice of
field functions and initial and boundary conditions the massless $j=1$
Weinberg-Tucker-Hammer equations contain all information that the Maxwell
equations for electromagnetic field have.  Thus, the former appear to be
of use in describing some physical processes for which that could be
necessitated or be convenient.
\end{abstract}


\baselineskip12pt

\section{Introduction}

In the sixties Joos~\cite{JOOS}, Weinberg~\cite{WEIN},  and  Weaver,
Hammer and Good~\cite{WEAV} proposed very attractive formalism (called
as the $2(2j+1)$ theory) for describing higher spin particles. For
instance,   as opposed to the Proca $4$-vector
potentials which transform according to the $(1/2,1/2)$ representation of
the Lorentz group, in the $j=1$ case
the ``bispinor"  functions are constructed via the
$(1,0)\oplus (0,1)$ representation what is on an equal footing to the
description of Dirac $j=1/2$ particles. The $2(2j+1)$- component
analogs of the Dirac functions in the momentum space were earlier defined
as
\begin{equation}\label{pos}
{\cal U}_\sigma ({\bf p})= {m\over \sqrt{2}} \left
(\matrix{ D^J \left (\alpha({\bf p})\right )\xi_\sigma\cr D^J \left
(\alpha^{-1\,\dagger}({\bf p})\right )\xi_\sigma\cr }\right )\quad,
\end{equation}
for positive-energy states, and
\begin{equation}\label{neg}
{\cal V}_\sigma ({\bf p})= {m\over \sqrt{2}} \left (\matrix{ D^J \left
(\alpha({\bf p})\Theta_{[1/2]}\right )\xi^*_\sigma\cr D^J \left (
\alpha^{-1\,\dagger}({\bf p}) \Theta_{[1/2]}\right
)(-1)^{2J}\xi^*_\sigma\cr }\right )\quad,
\end{equation} for
negative-energy states, e.g., ref.~\cite[p.107]{NOVO}. The following
notations is used
\begin{equation}
\alpha({\bf p})=\frac{p_0+m+(\bbox{\sigma}
\cdot{\bf p})}{\sqrt{2m(p_0+m)}},\quad
\Theta_{[1/2]}=-i\sigma_2\quad.
\end{equation}
For instance, in the case of spin $j=1$, one has
\begin{mathletters}
\begin{eqnarray}
&&D^{\,1}\left (\alpha({\bf p})\right ) \,=\,
1+\frac{({\bf J}\cdot{\bf p})}{m}+
\frac{({\bf J}\cdot{\bf p})^2}{m(p_0+m)}\quad,\\
&&D^{\,1}\left (\alpha^{-1\,\dagger}({\bf p})\right ) \,=\,
1-\frac{({\bf J}\cdot{\bf p})}{m}+
\frac{({\bf J}\cdot{\bf p})^2}{m(p_0+m)}\quad,  \\
&&D^{\,1}\left (\alpha({\bf p}) \Theta_{[1/2]}\right ) \,=\,
\left [1+\frac{({\bf J}\cdot{\bf p})}{m}+
\frac{({\bf J}\cdot{\bf p})^2}{m(p_0+m)}\right ]\Theta_{[1]}\quad, \\
&&D^{\,1}\left (\alpha^{-1\,\dagger}({\bf p}) \Theta_{[1/2]}\right ) \,=\,
\left [1-\frac{({\bf J}\cdot{\bf p})}{m}+
\frac{({\bf J}\cdot{\bf p})^2}{m(p_0+m)}\right ]\Theta_{[1]}\quad;
\end{eqnarray}
\end{mathletters}
$\Theta_{[1/2]}$,\,$\Theta_{[1]}$ are the Wigner time-reversal operators
for spin 1/2 and 1, respectively. These definitions lead to the
formulation in which the physical content given by
positive and negative-energy ``bispinors"
is the same (like in the papers of Weinberg and in the further
consideration of Tucker and Hammer~\cite{TUCK1}). In spite of the extensive
elaboration of the Weinberg $2(2j+1)$- component theory since the sixties,
{\it e.g.}, refs.~\cite{TUCK0,VLAS,GREI,SANT,LUKA,DVOE1,MISH} those
researches did not provide us new significant insights in the particle
physics.

Recently, a physically different construct in the
$(1,0)\oplus (0,1)$ representations has been proposed~\cite{AHLU1}. Its
remarkable feature is: a boson and its antiboson can possess opposite
intrinsic parities. The author of those papers wrote: ``\ldots purely by
accident, in an attempt to understand an old work of Weinberg~\cite{WEIN}
and to investigate the possible kinamatical origin for the violation of
$P$, $CP$, and other discrete symmetries~\cite{AHLU2}, a Wigner-type
quantum field theory~\cite{WIG2} was constructed for a spin-one
boson."\,\,\,\footnote{Let us note that some steps in this direction have
beed made earlier~\cite{SANK} but, unfortunately, the author of the papers
of 1965 did not realize all possible physical consequences following from
his equation.}\, The  definition of the negative-energy solutions in this
construct is similar to the Dirac construct for the spin-$1/2$ case:
\begin{equation}
{\cal V}_\sigma ({\bf p}) = \gamma_5 {\cal U}_\sigma ({\bf p}) =
(-1)^{1-\sigma} S^c_{[1]} {\cal U}_{-\sigma} ({\bf p})\quad,
\end{equation}
with $S^c_{[1]}$
being the charge conjugation matrix in the $(1,0)\oplus (0,1)$
representation~[13a,14]. They can be built by means of the same procedure
like used in Eqs. (\ref{pos}) and (\ref{neg}) but with taking into account
the possibility of an additional phase factor for up- (down-) components in
the bispinorial $j=1/2$ basis, see, e.g.,~\cite{AHLU2,DVOE2,DVOE3,DVOE3A}.

On the other hand, the interest in antisymmetric tensor
fields, {\it e.g.},~\cite{GURS,TAKA,HAYA,KALB,BOYA,AVDE,BIRM,CHIZ},
exists for a long time and even grows in connection with recent
discoveries of tensor couplings in the $\pi^-$ and $K^+$-meson decays.
These fields also should transform according to the $(1,0)\oplus (0,1)$
representation.

In the present paper we give a mapping between antisymmetric tensor
fields and Weinberg $j=1$ ``bispinors", hence propose the Lagrangian
formalism for a particular model in the $(1,0)\oplus (0,1)$ representation
and emphasize consequences relevant to the present situation in the
fundamental physics.  This paper comprises ideas presented
in~\cite{DVOE3,DVOE10,DVOE11,DVOE12,DVOE13,DVOE14}.

\section{Mapping between antisymmetric tensor and Weinberg formulations}

Let us begin with the Proca equations for a $j=1$ massive particle
\begin{eqnarray}\label{eq:01}
&&\partial_\mu F_{\mu\nu} = m^2 A_\nu \quad, \\
&&F_{\mu\nu} = \partial_\mu A_\nu - \partial_\nu A_\mu
\end{eqnarray}
in the form given by~\cite{SANK,LURI}.
The Euclidean metric,
$x_\mu =  (\vec x, x_4 =it)$ and notation
$\partial_\mu = (\vec \nabla, -i\partial/\partial t)$,
$\partial_\mu^2 = \vec \nabla^{\,2} -\partial_t^2$, are
used. By means of the choice of  $F_{\mu\nu}$ components
as the physical variables one can rewrite the set of equations to
\begin{equation}\label{eq:eq}
m^2 F_{\mu\nu} =\partial_\mu \partial_\alpha F_{\alpha\nu}
-\partial_\nu \partial_\alpha F_{\alpha\mu}
\end{equation}
and
\begin{equation}\label{eq:2}
\partial_\lambda^2 F_{\mu\nu} = m^2 F_{\mu\nu}\quad.
\end{equation}
It is easy to show that they can be represented in the form
($F_{44}=0$, $F_{4i} =i E_i$ and $F_{jk} =\epsilon_{jki} B_i$;\,
$p_\alpha=-i\partial_\alpha$):
\begin{eqnarray}\label{eq:aux1}
\cases{(m^2 +p_4^2) E_i +p_i p_j E_j +
i\epsilon_{ijk} p_4 p_j B_k=0& \cr
&\cr
(m^2 +\vec p^{\,2}) B_i -p_i p_j B_j +
i\epsilon_{ijk} p_4 p_j E_k =0\quad, &}
\end{eqnarray}
or
\begin{eqnarray}\label{eq:aux}
\cases{\left [ m^2 +p_4^2 +\vec p^{\,2} -
(\vec J \vec p)^2 \right ]_{ij}
E_j +p_4 (\vec J \vec p)_{ij} B_j = 0&\cr
\left [ m^2 +(\vec J \vec p)^2 \right ]_{ij} B_j +
p_4 (\vec J \vec p)_{ij} E_j =0\quad. &}
\end{eqnarray}
Adding and subtracting the  obtained equations yield
\begin{eqnarray}
\cases{m^2 (\vec E +i\vec B)_i + p_\alpha p_\alpha \vec E_i
- (\vec J \vec p)^2_{ij} (\vec E -i \vec B)_j
+ p_4 (\vec J \vec p)_{ij} (\vec B +i\vec E)_j = 0 &\cr
&\cr
m^2 (\vec E - i\vec B)_i + p_\alpha p_\alpha \vec E_i
- (\vec J\vec p)^2_{ij}
(\vec E +i\vec B)_j + p_4 (\vec J \vec p)_{ij}
(\vec B -i\vec E)_j = 0\quad, &}
\end{eqnarray}
with $(\vec J_i)_{jk} = -i\epsilon_{ijk}$ being
the $j=1$ spin matrices.
Equations are equivalent (within a constant factor) to
the Hammer-Tucker equation~\cite{TUCK1},
see also~\cite{DVOE1,VLAS}
\begin{equation}\label{eq:Tucker}
(\gamma_{\alpha\beta}p_\alpha p_\beta
+p_\alpha p_\alpha +2 m^2 ) \psi_1 =0 \quad ,
\end{equation}
in the case of the choice $\chi= \vec E +i\vec B$
and $\varphi =\vec E -i\vec B$,\,\,\,
$\psi_1 = \mbox{column} (\chi , \quad \varphi)$. Matrices
$\gamma_{\alpha\beta}$ are the
covariantly defined matrices of Barut,
Muzinich and Williams~\cite{BARU}.
The equation (\ref{eq:Tucker}) for massive particles is
characterized by positive- and negative-energy solutions with a physical
dispersion only $E_p =\pm \sqrt{\vec p^{\,2} + m^2}$, the determinant is
equal to
\begin{equation}
\mbox{Det} \left [\gamma_{\alpha\beta} p_\alpha p_\beta
+p_\alpha p_\alpha +2m^2 \right ] = -64 m^6 (p_0^2 -\vec p^{\,2}
-m^2)^3\quad, \label{det}
\end{equation}
but some points concerned with
a massless limit should be clarified properly.\footnote{Questions of
the correct relativistic dispersion relations of different $j=1$ equations
(both massive and massless) and of particle interpretations of these
solutions were also discussed in ref.~[35b]. For instance it was shown
that the Maxwell's equations possess `acausal' solution with the energy
$E=0$ and the Weinberg equation, while has common solutions with the
solutions of the Maxwell's equations, does {\it not} reduce entirely to
the set of Maxwell's equations in the massless limit.  The author of~[2b]
felt some unsatisfaction when discussed this question (see the first line
after Eqs.  (4.21,4.22) of ref.~[2b]) but he missed to indicate in a clear
manner that the matrix $(\vec J\cdot \vec p)$ has no the inverse one.
Several groups proposed recently interpretations of the $E=0$ solution.
One of them can be connected with the `action-at-a-distance' concept.  If
accept this viewpoint, the electromagnetic field has probably an
essentially non-local origins and it is connected with the structure of
space-time itself.} Following to the analysis of
ref.~[35b,p.1972]\footnote{I mean that some fraction of the operator
$\delta_{\alpha\beta} p_\alpha p_\beta$ acting on physically permittable
states can be substituted as $m^2 \leftrightarrow -\delta_{\alpha\beta}
p_\alpha p_\beta$.  The general equation can also be obtained by means of
setting up the generalized Ryder-Burgard
relation~\cite{AHLU1,DVOE3,DVOE3A}.} and in accordance with the Dirac
technique for obtaining wave equations~\cite{DIRAC} one can conclude that
other equations with the physical dispersion could be obtained from
\begin{equation}
(\gamma_{\alpha\beta}p_\alpha p_\beta +a p_\alpha p_\alpha +bm^2) \psi =0
\quad,
\end{equation}
with $a$ and $b$ being some numerical constants.
As a result of taking into account $E^2 -\vec
p^{\,2} = m^2$ we draw that the infinity number of equations with
the appropriate dispersion exists provided that $b$ and $a$ are connected
as follows:  $$\frac{b}{a+1}=1 \qquad \mbox{or} \qquad
\frac{b}{a-1}=1\quad.$$ However, there are only two equations which do not
have `acausal' solutions.  The second one (with $a=-1$ and $b=-2$)
is\footnote{
The determinant of the matrix in the left side of the following equation
is also given by the formula (\ref{det}).}
\begin{equation}
(\gamma_{\alpha\beta}p_\alpha p_\beta - p_\alpha p_\alpha - 2m^2) \psi_2
=0 \quad .  \label{eq:Tucker2}
\end{equation}
Thus, we have found the `double' of the
Hammer-Tucker equation.  In the tensor form it leads to the equations
which are dual  to (\ref{eq:aux1})
\begin{eqnarray}
\cases{(m^2 +\vec p^{\,2})C_i - p_i
p_j C_j - i\epsilon_{ijk} p_4 p_j D_k = 0 & \cr
&\cr
(m^2 +p_4^2 )D_i + p_i p_j
D_j - i\epsilon_{ijk} p_4 p_j C_k =0\quad. &}
\end{eqnarray}
They can be
rewritten in the form, {\it cf.} (\ref{eq:eq}),
\begin{equation}\label{eq:eqd}
m^2 \widetilde F_{\mu\nu} =\partial_\mu \partial_\alpha \widetilde
F_{\alpha\nu}
-\partial_\nu \partial_\alpha \widetilde F_{\alpha\mu} \quad ,
\end{equation}
with $\widetilde F_{4i} = iD_i$ and $\widetilde F_{jk} =
- \epsilon_{jki} C_i$.
The vector $C_i$ is an analog of $E_i$ and $D_i$ is an analog of $B_i$
because in some cases it is convenient to equate
$\widetilde F_{\mu\nu} ={1\over 2} \epsilon_{\mu\nu\rho\sigma}
F_{\rho\sigma}$, $\epsilon_{1234} = -i$.
The following
properties of the antisymmetric Levi-Civita tensor
$$ \epsilon_{ijk}
\epsilon_{ijl} =2\delta_{kl}\quad, \quad \epsilon_{ijk} \epsilon_{ilm} =
(\delta_{jl} \delta_{km} -\delta_{jm} \delta_{kl} )\quad,$$
and
$$ \epsilon_{ijk}
\epsilon_{lmn} =  \mbox{Det}\, \pmatrix{\delta_{il} & \delta_{im} &
\delta_{in}\cr
\delta_{jl} & \delta_{jm} & \delta_{jn}\cr
\delta_{kl} & \delta_{km} & \delta_{kn}} \quad $$
have been used.

Comparing the structure of the Weinberg equation ($a=0$, $b=1$)
with the Hammer-Tucker `doubles' one can convince ourselves
that the former can be represented in the tensor form:
\begin{equation}\label{eq:3}
m^2 F_{\mu\nu} =\partial_\mu \partial_\alpha F_{\alpha\nu}
-\partial_\nu \partial_\alpha F_{\alpha\mu} + {1\over 2}
(m^2 - \partial_\lambda^2) F_{\mu\nu}\quad,
\end{equation}
that corresponds to Eq. (\ref{w1}).
However, as we learnt, it is possible to build an equation --- `double'\,:
\begin{equation}\label{eq:4} m^2 \widetilde F_{\mu\nu} =\partial_\mu
\partial_\alpha \widetilde F_{\alpha\nu} -\partial_\nu \partial_\alpha
\widetilde F_{\alpha\mu} + {1\over 2} (m^2 -\partial_\lambda^2) \widetilde
F_{\mu\nu}\quad, \end{equation} that corresponds to Eq. (\ref{w2}).  The
Weinberg's set of equations is written in the form:
\begin{eqnarray}\label{w1}
(\gamma_{\alpha\beta} p_\alpha p_\beta + m^2 )\psi_1 &=& 0\quad,\\
\label{w2}
(\gamma_{\alpha\beta} p_\alpha p_\beta - m^2) \psi_2 &=& 0 \quad.
\end{eqnarray}
Thanks to the Klein-Gordon equation (\ref{eq:2}) these equations
are equivalent to the Proca tensor equations (and to
the Hammer-Tucker ones) in a free case.
However, if interaction is included, one cannot say that.
Thus, the general solution describing the $j=1$ states
can be presented as a superposition
\begin{equation}\label{eq:super}
\Psi^{(1)} = c_1 \psi_1^{(1)} + c_2 \psi_2^{(1)} \quad,
\end{equation}
where the constants $c_1$
and $c_2$ are to be defined from the boundary, initial
and normalization conditions.
Let me note a surprising fact:
while both the massive Proca equations (or the Hammer-Tucker ones)
and the Klein-Gordon equation do not possess ``non-physical"
solutions, their sum, Eqs. (\ref{eq:3},\ref{eq:4}), or
the Weinberg equations (\ref{w1},\ref{w2}), acquire tachyonic
solutions.
Next, equations (\ref{w1}) and (\ref{w2}) can
recast in another form (index $``T"$ denotes a transpose matrix):
\begin{eqnarray}\label{w11}
\left [\gamma_{44} p_4^2 +2\gamma_{4i}^T p_4 p_i +\gamma_{ij}p_i p_j
-m^2\right ] \psi_1^{(2)} &=&0 \quad ,\\ \label{w21}
\left [\gamma_{44} p_4^2 +2\gamma_{4i}^T p_4 p_i +\gamma_{ij}p_i p_j
+m^2 \right ] \psi_2^{(2)} &=&0 \quad ,
\end{eqnarray}
respectively, if understand
$\psi_1^{(2)} \sim \mbox{column} (B_i + iE_i ,\quad B_i -iE_i)
=i\gamma_5 \gamma_{44} \psi_1^{(1)}$ and
$\psi_2^{(2)} \sim \mbox{column} (D_i + i
C_i, \quad D_i - iC_i )= i\gamma_5 \gamma_{44} \psi_2^{(1)}$.
The general solution is again a linear combination
\begin{equation}
\Psi^{(2)} =c_1 \psi_1^{(2)} + c_2 \psi_2^{(2)}\quad.
\end{equation}

From, {\it e.g.}, Eq. (\ref{w1}),
dividing $\psi^{(1)}_1$
into longitudinal and transversal parts one can come to
the equations
\begin{eqnarray}
\lefteqn{\left [ E^2 -\vec p^{\,2}\right ](\vec E+i\vec B)^{\parallel}
-m^2 (\vec E-i\vec B)^{\parallel} +\nonumber}\\
&+&\left [ E^2 +\vec p^{\,2}- 2 E (\vec J\vec p)\right ]
(\vec E+i\vec B)^{\perp} - m^2 (\vec E-i\vec B)^{\perp} =0\quad,\label{cl1}
\end{eqnarray}
and
\begin{eqnarray}
\lefteqn{\left [ E^2 -\vec p^{\,2}\right ](\vec E-i\vec B)^{\parallel}
-m^2 (\vec E+i\vec B)^{\parallel} +\nonumber}\\
&+&\left [ E^2 +
\vec p^{\,2}+2 E (\vec J\vec p)\right ](\vec E-i\vec B)^{\perp} -
m^2 (\vec E+i\vec B)^{\perp} =0 \quad.\label{cl2}
\end{eqnarray}
One can see that in the classical field theory
antisymmetric tensor matter fields
are the fields with the transversal components
in massless limit. In this connection statements of the ``longitudinal
nature" of  the antisymmetric tensor field after quantization, made by
several authors~\cite{HAYA,KALB,AVDE} and~[28a], are very surprising. As a
matter of fact these authors contradicted with the Correspondence
Principle. We discuss this question below.

Under the transformations $\psi_1^{(1)}
\rightarrow \gamma_5 \psi_2^{(1)}$ or $\psi_1^{(2)}
\rightarrow \gamma_5 \psi_2^{(2)}$
the set of equations (\ref{w1}) and (\ref{w2}), or
(\ref{w11}) and (\ref{w21}), leaves to be invariant.
The origin of this fact is the dual invariance of the
set of the Proca equations.  In a matrix form
dual transformations correspond to the chiral transformations
(see for discussion, {\it e.g.}, ref.~\cite{STRA}).

Another equation has been proposed in refs.~\cite{SANK,AHLU1}
\begin{equation}\label{eq:Sankar}
(\gamma_{\alpha\beta} p_\alpha p_\beta +
\wp_{u,v} m^2 )\psi =0\quad,
\end{equation}
where $\wp_{u,v}=i(\partial/\partial t)/E$,
what distinguishes
$u$- (positive-energy) and $v$- (negative-energy) solutions.
For instance, in ~[13a,footnote 4] it is claimed that
\begin{equation}
\psi^+_\sigma (x) = \frac{1}{(2\pi)^3} \int
\frac{d^3 p}{2\omega_p} u_\sigma (\vec p) e^{ipx} \quad,
\end{equation}
$\omega_p =\sqrt{m^2 +\vec p^{\,2}}$, $p_\mu x_\mu
=\vec p \vec x -Et$, must be described by the equation
(\ref{w1}), in the meantime,
\begin{equation}
\psi^-_\sigma (x) =
\frac{1}{(2\pi)^3} \int \frac{d^3 p}{2\omega_p}
v_\sigma (\vec p) e^{-ipx}\quad,
\end{equation}
by the equation (\ref{w2}).  Nevertheless, calculating
the determinants (\ref{det}) of the equations
(\ref{eq:Tucker},\ref{eq:Tucker2})  we convinced ourselves that
the  first one has the {\it negative-energy} solutions and the second one,
{\it the positive-energy} solutions. The same is true for both Weinberg
equations, they are also have these solutions and  below we are going to
give their explicit forms.  The question of the choice of appropriate
equations for different physical systems was  discussed in
refs.~\cite{AHLU2,DVOE2,DVOE3}. The answer depends on desirable
particle properties with respect to discrete symmetries.

Let me consider the question of the `double' solutions
on the basis of spinorial analysis. In ref.~[16a,p.1305]
(see also~\cite[p.60-61]{LAND})
relations between the Weinberg $j=1$ ``bispinor" (bivector, indeed)
and symmetric spinors of $2j$- rank have been discussed.
It was noted there: ``The wave
function may be written in terms of two
three-component functions $\psi=\mbox{column}
(\chi \quad \varphi)$,
that, for the continuous group, transform independently
each of other and that are related to two symmetric
spinors:
\begin{eqnarray}
&&\chi_1 = \chi_{\dot 1\dot 1} , \quad \chi_2 = \sqrt{2}
\chi_{\dot 1 \dot 2} , \quad \chi_3 =
\chi_{\dot 2 \dot 2} \quad,\\
&& \varphi_1 = \varphi^{11} , \quad \varphi_2 = \sqrt{2}
\varphi^{12} , \quad \varphi_3 = \varphi^{22} \quad,
\end{eqnarray}
when the standard representation for the spin-one
matrices, with $S_3$ diagonal is used."
Under the inversion operation we
have the following rules~\cite[p.59]{LAND}:
$\varphi^\alpha \rightarrow \chi_{\dot\alpha}$,\,
$\chi_{\dot
\alpha} \rightarrow \varphi^{\alpha}$,\,
$\varphi_\alpha \rightarrow -\chi^{\dot \alpha}$
and $\chi^{\dot\alpha}\rightarrow -\varphi_\alpha$.
Hence, one can deduce (if one understand $\chi_{\dot\alpha\dot\beta}
=\chi_{\{\dot\alpha} \chi_{\dot\beta\}}$\, ,\, $\varphi^{\alpha\beta}
=\varphi^{\{\alpha}\varphi^{\beta\}}$)
\begin{eqnarray}
&&\chi_{\dot 1 \dot 1} \rightarrow \varphi^{11} \quad, \quad
\chi_{\dot 2 \dot 2} \rightarrow \varphi^{22} \quad, \quad
\chi_{ \{ \dot 1 \dot 2 \} } \rightarrow
\varphi^{ \{ 12 \} } \quad,\\
&& \varphi^{11} \rightarrow \chi_{\dot 1 \dot 1} \quad, \quad
\varphi^{22} \rightarrow \chi_{\dot 2 \dot 2} \quad, \quad
\varphi^{ \{ 12 \} } \rightarrow \chi_{ \{ \dot 1 \dot 2 \} } \quad.
\end{eqnarray}
However, this definition of symmetric spinors
of the second rank $\chi$ and $\varphi$ is ambiguous.
We are also able to define, {\it e.g.}, $\tilde \chi_{\dot\alpha\dot\beta}
=\chi_{\{\dot\alpha} H_{\dot\beta\}}$ and
$\tilde \varphi^{\alpha\beta} = \varphi^{\{\alpha}
\Phi^{\beta\}}$,
where $H_{\dot\beta} = \varphi_{\beta}^{*}$,
$\Phi^{\beta} =(\chi^{\dot\beta})^{*}$.
It is straightforward to show that in the framework
of the second definition we
have under the space-inversion operation:
\begin{eqnarray}
&&\tilde \chi_{\dot 1 \dot 1} \rightarrow
-\tilde \varphi^{11} \quad , \quad
\tilde \chi_{\dot 2 \dot 2} \rightarrow
- \tilde \varphi^{22} \quad , \quad
\tilde \chi_{ \{ \dot 1 \dot 2 \} } \rightarrow
- \tilde \varphi^{ \{ 12 \} } \quad, \\
&&\tilde\varphi^{11} \rightarrow -\tilde \chi_{\dot 1 \dot 1}\quad , \quad
\tilde \varphi^{22} \rightarrow -\tilde \chi_{\dot 2
\dot 2}\quad , \quad
\tilde \varphi^{ \{ 12 \} } \rightarrow
-\tilde \chi_{ \{ \dot 1 \dot 2 \} } \quad .
\end{eqnarray}
The Weinberg ``bispinor"
$(\chi_{\dot\alpha\dot\beta} \quad \varphi^{\alpha\beta})$
corresponds to the equations (\ref{w11}) and (\ref{w21}) ,
meanwhile
$(\tilde \chi_{\dot\alpha\dot\beta}\quad
\tilde\varphi^{\alpha\beta})$, to
the equations (\ref{w1}) and (\ref{w2}).
Similar conclusions can be achieved in
the case of the parity definition
as $P^2 = -1$. Transformation rules are then
$\varphi^\alpha \rightarrow i\chi_{\dot\alpha}$, $\chi_{\dot\alpha}
\rightarrow i\varphi^\alpha$,
$\varphi_\alpha \rightarrow -i\chi^{\dot\alpha}$
and $\chi^{\dot\alpha}\rightarrow -i\varphi_\alpha$,
ref.~\cite[p.59]{LAND} .
Hence, $\chi_{\dot\alpha\dot\beta} \leftrightarrow
-\varphi^{\alpha\beta}$
and $\tilde \chi_{\dot\alpha\dot\beta} \leftrightarrow
- \tilde\varphi^{\alpha\beta}$, but $\varphi_{\alpha}^{\quad\beta}
\leftrightarrow \chi^{\alpha}_{\quad\beta}$ and $\tilde
\varphi_{\alpha}^{\quad\beta} \leftrightarrow
\tilde \chi^{\alpha}_{\quad\beta}$\quad.

Next, in the previous formulations of the Weinberg theory the following
Lagrangian was proposed~\cite{GREI,SANT}  and~[11b,28a,b]:
\begin{equation}\label{eq:Lagra}
{\cal L}^{W}=-\partial_{\mu}\overline\psi\gamma_{\mu\nu}
\partial_\nu\psi-m^2\overline\psi\psi \quad,
\end{equation}
$\gamma_{\mu\nu}$ are the Barut-Muzinich-Williams matrices
which are chosen to be Hermitian. It is scalar, {\it cf.}~[28a],
and Hermitian\footnote{When the Euclidean metric is used the only
inconvenience must be taken in mind where it is necessary: we need imply
that $\partial_\mu^\dagger = (\vec \nabla, - \partial /\partial x_4)$,
provided that $\partial_\mu = (\vec \nabla, \partial /\partial  x_4)$,
ref.~\cite{LURI}.}, {\it cf.}~\cite{GREI} and it contains only
first-order time derivatives.
Again implying
interpretation of the ``6-spinor"  as\footnote{One can also choose
$$\psi^{(2)}= \pmatrix{\vec E+i\vec B\cr -\vec E+i\vec B} =\gamma_5
\psi\quad.$$
Since $\overline \psi^{(2)} =-\overline \psi \gamma_5$ the dynamical
term (\ref{eq:Lagra}) is not changed. But the sign in the mass
term would be inverse.}
\begin{eqnarray}\label{eq:EH}
\cases{ \chi=\vec E+i\vec
B\quad,& $ $\cr \phi=\vec E-i\vec B\quad,& $ $}
\end{eqnarray}
$\psi =\mbox{column}
(\chi  \quad \phi)$,  $\vec E$ and $\vec B$ are the real 3-vectors,
the  Lagrangian (\ref{eq:Lagra}) can be re-written in the  following way:
\begin{equation}\label{eq:Lagran}
{\cal L}^{AT}=-(\partial_\mu F_{\nu\alpha})(\partial_\mu
F_{\nu\alpha}) +
2(\partial_\mu F_{\mu\alpha})(\partial_\nu F_{\nu\alpha})
+ 2(\partial_\mu F_{\nu\alpha})(\partial_\nu F_{\mu\alpha})
+m^2 F_{\mu\nu} F_{\mu\nu}\quad.
\end{equation}
In  a massless limit this form of the Lagrangian leads to the
Euler-Lagrange equation
\begin{equation} ({\,\lower0.9pt\vbox{\hrule
\hbox{\vrule height 0.2 cm \hskip 0.2 cm \vrule height 0.2cm}\hrule}\,}
-m^2) F_{\alpha\beta}-2(\partial_{\beta}F_{\alpha\mu,\mu}-
\partial_{\alpha}F_{\beta\mu,\mu})=0 \quad,
\end{equation}
where ${\,\lower0.9pt\vbox{\hrule \hbox{\vrule height 0.2 cm
\hskip 0.2 cm
\vrule height 0.2 cm}\hrule}\,}=\partial_{\nu}\partial_{\nu}$.
After the application of the generalized Lorentz condition~\cite{HAYA} the
massless Lagrangian (\ref{eq:Lagran}) becomes to be  equivalent to the
Lagrangian of  a free massless skew-symmetric field given in
ref.~\cite{HAYA}:
\begin{equation}\label{eq:LagHa}
{\cal L}^{H}=\frac{1}{8}F_k F_k \quad,
\end{equation}
with $F_k=i\epsilon_{kjmn}F_{jm,n}$. It is re-written in ($m=0$):
\begin{eqnarray}
{\cal L}^{H}&=&-{1\over 4}(\partial_{\mu} F_{\nu\alpha})
(\partial_{\mu} F_{\nu\alpha})+
{1\over 2} (\partial_{\mu} F_{\nu\alpha})(\partial_{\nu}
F_{\mu\alpha})=\nonumber\\
&=&{1 \over 4}{\cal L}^{AT}-{1\over 2}(\partial_{\mu}
F_{\mu\alpha})(\partial_{\nu} F_{\nu\alpha})\quad,
\end{eqnarray}
what proves the statement made above.
After the  application of the Fermi method {\it mutatis mutandis}
as in ref.~\cite{HAYA} ({\it cf.} with the quantization procedure for
a 4-vector potential field)
one achieved  the  result  that  the Lagrangians
(\ref{eq:Lagra}) and (\ref{eq:LagHa})
describe  massless particles
possessing longitudinal physical components only. Transversal
components are removed by means of the ``gauge" transformation
\begin{equation}\label{eq:gauge}
F_{\mu\nu}\rightarrow
F_{\mu\nu}+A_{[\mu\nu]}=F_{\mu\nu}+\partial_{\nu}
\Lambda_{\mu}-
\partial_{\mu}\Lambda_{\nu}
\end{equation}
(or by  the  transformation similar to the above but applied to the Weinberg
bivector). This is a contradiction to which has been paid attention
in~[28a,b]: the $j=1$ antisymmetric tensor field was believed to possess
the longitudinal component only, the helicity  is therefore equal to
$\lambda=0$. In the meantime, they transform
according to  the  $(1,0) +(0,1)$ representation of the Lorentz group
(like a Helmoltz-Weinberg bivector).  How is the Weinberg theorem,
ref.~\cite{WEIN},  for the $(A,B)$ representation to be treated in this
case?\,\footnote{Let me recall that the Weinberg theorem states: {\it The
fields constructed from the massless particle operator \,\,$a (\vec
p,\lambda)$\,\, of the definite helicity transform according to
representation\,\, $(A,B)$ \,\,such that \,\,$B-A=\lambda$.}} If we
want to have well-defined creation and annihilation operators the
antisymmetric tensor field should have helicities $\lambda =\pm
1$.\footnote{Several authors indicated this from different viewpoints in
refs.~\cite{GURS,TAKA,BOYA,AHLU1}.}
Moreover, do the claims of the ``longitudinal nature" of the antisymmetric
tensor field and, hence, the Weinberg $j=1$ field signify that we must
abandon the Correspondence Principle:  in the classical physics we know
that an antisymmetric tensor field is with transversal components, see
also Eqs. (\ref{cl1},\ref{cl2})?

This contradiction has been analyzed in
refs.~\cite{DVOE11,DVOE12,DVOE13,DVOE14,EVAN,DVOE96} in detail. The
result achieved is:  transversal components are always
linked with longitudinal spin components and can be decoupled only in
particular cases. Using the Weinberg formalism we provide
additional support to this conclusion in the following Section.

We conclude this Section: both the theory of Ahluwalia {\it et al.}
~\cite{AHLU1,AHLU2} and the model based on the use of $\psi_1$ and
$\psi_2$ are connected with the antisymmetric tensor matter field
description.  They have to be quantized consistently. Special attention
should be paid to the translational and rotational invariance (the
conservation of energy-momentum and angular momentum, indeed), the
interaction representation, causality, locality and covariance of
the theory, {\it i.e.} to all topics, which are the axioms of the modern
quantum field theory ~\cite{ITZY,BOGO}. A consistent theory has also to
take into account the degeneracy of states:  two dual functions $\psi_1$
and $\psi_2$ (or $F_{\mu\nu}$ and $\widetilde F_{\mu\nu}$, the `doubles')
are considered to yield the same spectrum.

\section{What particles are described by the Weinberg theory?}

In  the previous Section the concept of the Weinberg $j=1$ field as a
system of degenerate states has been proposed. As a matter of fact
a model with the Weinberg `doubles'  is equivalent to dual electrodynamics
with the antisymmetric tensor field $F_{\mu\nu}$ and its dual $\widetilde
F_{\mu\nu}$. Unfortunately, many works concerned with the dual
theories~\cite{BOYA,STRA,CABB,RECA} did not worked out quantization
issues in detail and many specific features of such a consideration have
not been taken into account earlier.\footnote{We wish to mention that dual
formulations of the Dirac field, the $(1/2, 0)\oplus (0, 1/2)$
representation, have also been considered, {\it e.  g.},
ref.~\cite{MARK,BRAN,ZIINO,DVOE2,DVOE3}.  The interaction of the
Dirac field with the dual fields $F_{\mu\nu}$ and $\widetilde F_{\mu\nu}$
has been considered in ref.~\cite{LAVR} (this implies the existence of the
anomalous electric dipole moment of a fermion).}

We begin with the  Lagrangian  which is similar
to Eq. (\ref{eq:Lagra}) but includes additional terms
which respond to the Weinberg `double'. Here it
is:\footnote{Of course, one can use another form with substitutions:
$\psi_{1,2}^{(1)} \rightarrow \psi_{2,1}^{(2)}$ and $\gamma_{\mu\nu}
\rightarrow \widetilde \gamma_{\mu\nu}$, where $\widetilde \gamma_{\mu\nu}
\equiv \gamma_{\mu\nu}^T \equiv \gamma_{44} \gamma_{\mu\nu} \gamma_{44}$.}
\begin{equation}\label{eq:Lagran1}
{\cal L}^{(1)} = -\partial_\mu \overline \psi_1  \gamma_{\mu\nu}
\partial_\nu \psi_1
-\partial_\mu \overline \psi_2 \gamma_{\mu\nu}
\partial_\nu \psi_2 - m^2 \overline \psi_1 \psi_1
+ m^2 \overline \psi_2 \psi_2 \quad .
\end{equation}
The Lagrangian (\ref{eq:Lagran1}) leads to the equations
(\ref{w1},\ref{w2})
which possess solutions with  a ``correct" bradyon physical
dispersion and  tachyonic solutions as well.  The second equation
coincides with the Ahluwalia {\it et al.} equation  for $v$ spinors (Eq.
(13) ref.~[13a]) or with Eq. (12) of ref.~[16c].  If
accept the concept of the Weinberg field as a set of degenerate states,
one has to allow for possible transitions $\psi_1 \leftrightarrow \psi_2$
(or  $F_{\mu\nu} \leftrightarrow \widetilde F_{\mu\nu}$).
From the first sight, one can
propose the Lagrangian with the following dynamical part:
\begin{equation}
{\cal L}^{(2^\prime)} = - \partial_\mu \overline \psi_1 \gamma_{\mu\nu}
\partial_\nu \psi_2 -\partial_\mu \overline \psi_2 \gamma_{\mu\nu}
\partial_\nu \psi_1  \quad, \nonumber
\end{equation}
where $\psi_1$ and $\psi_2$ are defined by the equations
(\ref{w1},\ref{w2}).  But, this form appears
not to  admit a mass term in a
usual manner.  From a mathematical viewpoint one can
find solution:  set $m^2$ to be pure imaginary quantity (or in the operator
formulation, the anti-Hermitian operator). We touched this case
earlier~\cite{DVOE12}. More logical approach seems to be in
regarding all four states described by Eqs.
(\ref{w1},\ref{w2},\ref{w11},\ref{w21}).
The following Lagrangian can be proposed in this case:
\begin{eqnarray}
{\cal L}^{(2)} &=& - \partial_\mu  \psi_1^{(1)\,\dagger} \widetilde
\gamma_{\mu\nu} \partial_\nu \psi_2^{(2)} - \partial_\mu
\psi_2^{(2)\,\dagger} \gamma_{\mu\nu} \partial_\nu \psi_1^{(1)}   -
\partial_\mu \psi_2^{(1)\,\dagger} \widetilde \gamma_{\mu\nu} \partial_\nu
\psi_1^{(2)} - \partial_\mu \psi_1^{(2)\,\dagger} \gamma_{\mu\nu}
\partial_\nu \psi_2^{(1)} -\nonumber\\ &-& m^2 \psi_2^{(2)\,\dagger}
\psi_1^{(1)} - m^2  \psi_1^{(1)\,\dagger} \psi_2^{(2)} + m^2
\psi_2^{(1)\,\dagger} \psi_1^{(2)} + m^2 \psi_1^{(2)\,\dagger}
\psi_2^{(1)}
\quad.  \label{eq:Lagran2} \end{eqnarray}
Both the Lagrangian (\ref{eq:Lagran1}) and (\ref{eq:Lagran2})
are scalars,\footnote{It is easy to verify this by means of taking into
account proposed interpretations of $\psi_i^{(k)} (x)$ which are connected
with the tensor $F^{\mu\nu}$ and its dual. There is also another way,
on the basis of the use of explicit forms of momentum-space
``6-spinors", see below.}
Hermitian and they contain only first-order time derivatives.  The both
lead to similar equations for $\psi_1^{(1,2)} (x)$ and $\psi_2^{(1,2)}
(x)$ but one should not forget about the difference in signs in mass terms
when considering the equations for $\psi_i^{(k)} (x)$.

At this point  I would like to regard the question of solutions
in the momentum space. Using the plane-wave expansion\footnote{I
stress that to keep a mathematical approach as general as
possible is in the aims of the present investigation. The relevance
of different photon spin states to different forms of field operators will
be studied in more detail in forthcoming publications.}
\begin{eqnarray}\label{pl1}
\psi_1^{(k)} (x) &=&\sum_\sigma \int \frac{d^3 p}{(2\pi)^3} \frac{1}{m
\sqrt{2E_p}} \left [ {\cal U}_1^{(k)\,\sigma} (\vec p) a^{(k)}_\sigma
(\vec p) e^{ipx} +{\cal V}_1^{(k)\,\sigma} (\vec p)
b^{(k)\,\dagger}_\sigma (\vec p) e^{-ipx} \right ]\quad,\\ \label{pl2}
\psi_2^{(k)} (x) &=&\sum_\sigma \int \frac{d^3 p}{(2\pi)^3}
\frac{1}{m\sqrt{2E_p}} \left [ {\cal U}_2^{(k)\,\sigma} (\vec p)
c^{(k)}_\sigma (\vec p) e^{ipx} +{\cal V}_2^{(k)\,\sigma} (\vec p)
d^{(k)\,\dagger}_\sigma (\vec p) e^{-ipx} \right ] \quad,
\end{eqnarray}
$E_p=\sqrt{\vec p^{\,2} +m^2}$, one can see that the momentum-space
`double' equations \begin{eqnarray}\label{eq:sp1} \left [ - \gamma_{44}
E^2 +2iE\gamma_{4i} \vec p_i +\gamma_{ij} \vec p_i \vec p_j  + m^2\right ]
{\cal U}_1^\sigma (\vec p) &=& 0 \quad (\mbox{or}\,\,{\cal V}_1^\sigma
(\vec p)) \quad, \\ \label{eq:sp2} \left [ - \gamma_{44} E^2
+2iE\gamma_{4i} \vec p_i +\gamma_{ij} \vec p_i \vec p_j  - m^2\right ]
{\cal U}_2^\sigma (\vec p) &=& 0 \quad (\mbox{or} \,\,{\cal V}_2^\sigma
(\vec p))\quad \end{eqnarray}
are satisfied by ``bispinors"
\begin{eqnarray}\label{bb1}
{\cal U}_1^{(1)\,\sigma} (\vec p)=
\frac{m}{\sqrt{2}}\pmatrix{\left [ 1+ {(\vec J\vec p)\over  m} +{(\vec J
\vec p)^2 \over  m(E+m)}\right ]\xi_\sigma \cr \left [ 1  - {(\vec J\vec
p)\over  m} +{(\vec J \vec p)^2 \over m(E+m)}\right ] \xi_\sigma\cr}\quad,
\end{eqnarray}
and
\begin{eqnarray}\label{bb2}
{\cal U}_2^{(1)\,\sigma} (\vec p)
=\frac{m}{\sqrt{2}}\pmatrix{\left [ 1+
{(\vec J\vec p)\over  m} + {(\vec J \vec p)^2 \over  m(E+m)}\right ]
\xi_{\sigma} \cr
\left [ - 1  + {(\vec J\vec p)\over  m} - {(\vec J \vec p)^2
\over  m(E+ m)}\right ] \xi_{\sigma}\cr}\quad,
\end{eqnarray}
respectively.
The form (\ref{bb1}) has been presented by Hammer, Tucker and
Novozhilov in refs.~\cite{TUCK1,NOVO}, see also~\cite{DVOE1}.  The
bispinor normalization in the cited papers is chosen to unit. However, as
mentioned in ref.~\cite{AHLU1} it is more convenient to work with
bispinors normalized to the mass, {\it e.g.}, $\pm m^{2j}$ in order to
make zero-momentum spinors to vanish in the massless limit.  Here and
below I keep the normalization of bispinors as in
ref.~\cite{AHLU1}. Bispinors of Ahluwalia {\it et al.},
ref.~\cite{AHLU1}, can be written in the  more compact  form:
\begin{eqnarray}
u^{\sigma}_{AJG} (\vec p) =
\pmatrix{\left [ m +\frac{(\vec J \vec p)^2}{E+m} \right ] \xi_{\sigma}\cr
(\vec J \vec p) \xi_{\sigma}\cr}\quad, \quad v^{\sigma}_{AJG} (\vec p)
=\pmatrix{0 & 1\cr 1 & 0\cr} u^{\sigma}_{AJG} (\vec p)\quad.
\end{eqnarray}
They coincide with the Hammer-Tucker-Novozhilov bispinors within a
normalization and a unitary transformation by ${\cal U}$ matrix:
\begin{eqnarray}
u^{\sigma}_{~\cite{AHLU1}} (\vec p) = m\,\,\, \cdot  {\bf U}
{\cal U}^{\sigma}_{~\cite{TUCK1,NOVO}} (\vec p) =
\frac{m}{\sqrt{2}}\pmatrix{1 & 1 \cr 1 & -1\cr}
{\cal U}^{\sigma}_{~\cite{TUCK1,NOVO}} (\vec p)\quad,
\end{eqnarray}
\begin{eqnarray}
v^{\sigma}_{~\cite{AHLU1}} (\vec p) = m\,\,\, \cdot {\bf U}
\gamma_{5} {\cal U}^{\sigma}_{~\cite{TUCK1,NOVO}} (\vec p) =
\frac{m}{\sqrt{2}}\pmatrix{1 & 1 \cr 1 & -1\cr} \gamma_{5}
{\cal U}^{\sigma}_{~\cite{TUCK1,NOVO}} (\vec p) \quad.
\end{eqnarray}
But, as we found the Weinberg equations
(with $+m^2$ and with $-m^2$)  have solutions with both positive- and
negative-energies. We have to propose the interpretation of the latter.
In the framework of this paper one can consider that
${\cal V}^{(1,2)}_\sigma (\vec p) = (-1)^{1-\sigma}\gamma_5 S^c_{[1]}
{\cal U}^{(1,2)}_{-\sigma} (\vec p)$ and, thus, the explicit form of the
negative-energy solutions would be the same as of the positive-energy
solutions in accordance with definitions (\ref{pos},\ref{neg}), see the
discussion in the Section I.  Thus, in
the case of a choice ${\cal U}_1^{(1)\,\sigma} (\vec p)$ and
${\cal V}_2^{(1)\,\sigma} \sim \gamma_5 {\cal U}_1^{(1)\,\sigma} (\vec p)$
as physical bispinors we come to the Bargmann-Wightman-Wigner-type (BWW)
quantum field model proposed by Ah\-lu\-wa\-lia {\it et al.} Of course,
following  to the same logic one can choose ${\cal U}_2^{(1)\,\sigma}$ and
${\cal V}_1^{(1)\,\sigma}$ bispinors and come to yet another version of
the BWW theory.  While in this case parities of a boson and its antiboson
are opposite, we have $-1$ for ${\cal U}-$ bispinor and $+1$ for ${\cal
V}-$ bispinor, {\it i.e.} different in the sign from the model of
Ahluwalia {\it et al.}.\footnote{At the present level of our knowledge
this mathematical difference has no physical significance, but we want to
stay in the most general frameworks and, perhaps, some forms of
interactions can lead to the observed physical difference between these
models.} In the meantime, the construct
proposed by Weinberg~\cite{WEIN} and developed in this paper is also
possible. I do not agree with the claim of the authors of
ref.~[13a,footnote 4] which states ${\cal V}^{(1)\,\sigma}_1 (\vec p)$ are
not solutions of the equation (\ref{w1}).  The origin of the possibility
that the ${\cal U}_i$- and ${\cal V}_i$- bispinors  in Eqs.
(\ref{eq:sp1},\ref{eq:sp2}) can coincide each other
is the following:  the Weinberg equations are of
the second order in  time derivatives.  The Bargmann-Wightman-Wigner
construct presented by Ahluwalia~\cite{AHLU1} is not the only construct in
the $(1,0)\oplus (0,1)$ representation and one can start with the earlier
definitions of the $2(2j+1)$ bispinors.

Next, in the Section II we gave two additional equations
(\ref{w11},\ref{w21}). Their solutions can  also be useful because
of the possibility of the use of the  Lagrangian form (\ref{eq:Lagran2}).
The solutions in the momentum representation are written as follows
\begin{eqnarray}\label{b11}
{\cal U}_1^{(2)\,\sigma} (\vec p)=
\frac{m}{\sqrt{2}}\pmatrix{\left [1- {(\vec J\vec p)\over m}
+{(\vec J \vec p)^2 \over m (E+m)}\right ]\xi_\sigma \cr \left [ -1  -
{(\vec J\vec p)\over m}  -{(\vec J \vec p)^2 \over  m (E + m)}\right ]
\xi_\sigma}\quad,
\end{eqnarray}
\begin{eqnarray}\label{b21}
{\cal U}_2^{(2)\,\sigma} (\vec p)=
=\frac{m}{\sqrt{2}}\pmatrix{\left [1-
{(\vec J\vec p) \over m} + {(\vec J \vec p)^2 \over m (E + m)}\right
]\xi_\sigma \cr \left [ 1  + {(\vec J\vec p) \over m} + {(\vec J \vec
p)^2 \over  m (E + m)}\right ] \xi_\sigma}\quad.
\end{eqnarray}
Therefore,  one has
${\cal U}_2^{(1)} (\vec p) = \gamma_5 {\cal U}_1^{(1)} (\vec p)$ and
$\overline {\cal U}_2^{(1)} (\vec p) = -\overline {\cal U}_1^{(1)} (\vec
p)\gamma_5$; \, ${\cal U}_1^{(2)} (\vec p) = \gamma_5\gamma_{44}
{\cal U}_1^{(1)} (\vec p)$ and $\overline {\cal U}_1^{(2)} = \overline
{\cal U}_1^{(1)} \gamma_5\gamma_{44}$; ${\cal U}_2^{(2)} (\vec p) =
\gamma_{44} {\cal U}_1^{(1)} (\vec p)$ and $\overline {\cal U}_2^{(2)}
(\vec p) =  \overline {\cal U}_1^{(1)}\gamma_{44}$.  In fact, they are
connected by transformations of the inversion group.

Let me now repeat the quantization procedure for
antisymmetric tensor field presented, {\it e.g.}, in ref.~\cite{HAYA},
however, it will be applied to the Weinberg field.
Let me trace contributions of ${\cal L}^{(1)}$ to
dynamical invariants.  From the definitions~\cite{LURI}:
\begin{eqnarray}
{\cal T}_{\mu\nu} &=& -\sum_i \left \{
\frac{\partial {\cal L}}{\partial (\partial_\mu \phi_i )}
\partial_\nu \phi_i
+ \partial_\nu \overline \phi_i\frac{\partial {\cal L}}{\partial
(\partial_\mu \overline \phi_i )} \right \}
+{\cal L}\delta_{\mu\nu}\quad,\\
P_\mu &=& \int {\cal P}_{\mu} (x) d^3 x = -i\int {\cal T}_{4\mu} d^3 x
\end{eqnarray}
one can find  the energy-momentum tensor\footnote{Finding the
classical dynamical invariants from the Lagrangian ${\cal L}^{(2)}$  does
not present any difficulties. Here they are:
\begin{eqnarray}
\lefteqn{{\cal T}_{\mu\nu}^{(2)} = \partial_\alpha \psi_1^{(1)\,\dagger}
\widetilde\gamma_{\alpha\mu} \partial_\nu  \psi_2^{(2)} + \partial_\alpha
\psi_1^{(2)\,\dagger}\gamma_{\alpha\mu} \partial_\nu \psi_2^{(1)} +
\partial_\alpha \psi_2^{(1)\,\dagger} \widetilde\gamma_{\alpha\mu}
\partial_\nu \psi_1^{(2)} + \partial_\alpha \psi_2^{(2)\,\dagger}
\gamma_{\alpha\mu}\partial_\nu \psi_1^{(1)} +}\\
&+&
\partial_\nu \psi_1^{(1)\,\dagger}
\widetilde\gamma_{\mu\alpha} \partial_\alpha  \psi_2^{(2)} +
\partial_\nu \psi_1^{(2)\,\dagger}\gamma_{\mu\alpha} \partial_\alpha
\psi_2^{(1)} + \partial_\nu \psi_2^{(1)\,\dagger}
\widetilde\gamma_{\mu\alpha} \partial_\alpha \psi_1^{(2)} +
\partial_\nu \psi_2^{(2)\,\dagger} \gamma_{\mu\alpha}\partial_\alpha
\psi_1^{(1)} + {\cal L}^{(2)} \delta_{\mu\nu}\quad\nonumber;
\end{eqnarray}
\begin{eqnarray}\label{H2}
{\cal H}^{(2)} &=& \int \left [ -\partial_4
\psi_1^{(1)\,\dagger} \gamma_{4 4} \partial_4 \psi_2^{(2)} +
\partial_i  \psi_1^{(1)\,\dagger}  \gamma_{ij} \partial_j
\psi_2^{(2)} - \partial_4 \psi_1^{(2)\,\dagger} \gamma_{4 4} \partial_4
\psi_2^{(1)} + \partial_i \psi_1^{(2)\,\dagger}
\gamma_{ij} \partial_j \psi_2^{(1)}-\right.\nonumber\\
&-& \left.
\partial_4  \psi_2^{(1)\,\dagger} \gamma_{4 4} \partial_4
\psi_1^{(2)} + \partial_i  \psi_2^{(1)\,\dagger} \gamma_{ij}
\partial_j \psi_1^{(2)} -
\partial_4  \psi_2^{(2)\,\dagger} \gamma_{4 4} \partial_4
\psi_1^{(1)} + \partial_i  \psi_2^{(2)\,\dagger} \gamma_{ij}
\partial_j \psi_1^{(1)} +\right.\nonumber\\
&+& \left. m^2  \psi_1^{(1)\,\dagger} \psi_2^{(2)} -
m^2 \psi_1^{(2)\,\dagger} \psi_2^{(1)} - m^2 \psi_2^{(1)\,\dagger}
\psi_1^{(2)} + m^2 \psi_2^{(2)\,\dagger} \psi_1^{(1)}\right ]
d^3 x \quad.
\end{eqnarray}
The charge operator and the spin tensor are
\begin{eqnarray}\label{qq2}
\lefteqn{{\cal J}^{(2)}_\mu = i \left
[\partial_\alpha \psi_1^{(1)\,\dagger}\widetilde \gamma_{\alpha\mu}
\psi_2^{(2)} + \partial_\alpha \psi_1^{(2)\,\dagger} \gamma_{\alpha\mu}
\psi_2^{(1)} + \partial_\alpha \psi_2^{(1)\,\dagger}
\widetilde\gamma_{\alpha\mu} \psi_1^{(2)} + \partial_\alpha
\psi_2^{(2)\,\dagger} \gamma_{\alpha\mu} \psi_1^{(1)}-\right .\nonumber}\\
&-& \left .
\psi_1^{(1)\,\dagger}\widetilde \gamma_{\mu\alpha}
\partial_\alpha \psi_2^{(2)} - \psi_1^{(2)\,\dagger}
\gamma_{\mu\alpha} \partial_\alpha \psi_2^{(1)} -
\psi_2^{(1)\,\dagger} \widetilde\gamma_{\mu\alpha}
\partial_\alpha\psi_1^{(2)} - \psi_2^{(2)\,\dagger}
\gamma_{\mu\alpha} \partial_\alpha\psi_1^{(1)}\right ]\quad;
\end{eqnarray}
\begin{eqnarray}\label{ss2}
\lefteqn{S_{\mu\nu,\lambda}^{(2)}  = i \left [ \partial_\alpha
\psi_1^{(1)\,\dagger} \widetilde\gamma_{\alpha\lambda}
N_{\mu\nu}^{\psi_2^{(2)}} \psi_2^{(2)} + \partial_\alpha
\psi_1^{(2)\,\dagger}\gamma_{\alpha\lambda}
N_{\mu\nu}^{\psi_2^{(1)}} \psi_2^{(1)} +\right.}\\
&&\left.\qquad\qquad\qquad\qquad +
\partial_\alpha \psi_2^{(1)\,\dagger} \widetilde\gamma_{\alpha\lambda}
N_{\mu\nu}^{\psi_1^{(2)}} \psi_1^{(2)} + \partial_\alpha
\psi_2^{(2)\,\dagger} \gamma_{\alpha\lambda} N_{\mu\nu}^{\psi_1^{(1)}}
\psi_1^{(1)} + \right.\nonumber\\
&+& \left. \psi_1^{(1)\,\dagger}
N_{\mu\nu}^{\psi_1^{(1)\,\dagger}} \widetilde\gamma_{\lambda\alpha}
\partial_\alpha \psi_2^{(2)} + \psi_1^{(2)\,\dagger}
N_{\mu\nu}^{\psi_1^{(2)\,\dagger}}\gamma_{\lambda\alpha}
\partial_\alpha \psi_2^{(1)} +
\psi_2^{(1)\,\dagger}
N_{\mu\nu}^{\psi_2^{(1)\,\dagger}} \widetilde\gamma_{\lambda\alpha}
\partial_\alpha \psi_1^{(2)} + \psi_2^{(2)\,\dagger}
N_{\mu\nu}^{\psi_2^{(2)\,\dagger}} \gamma_{\lambda\alpha}
\partial_\alpha\psi_1^{(1)} \right ]\quad.\nonumber
\end{eqnarray}
Questions of the translational invariance, the choice of
bispinors answering the physical states, the renormalizability of the
theory based on the ${\cal L}^{(2)}$, the possibility of existence of the
chiral charge for this system (like for the Majorana states in the
$(1/2,0)\oplus (0,1/2)$ representation, what has been shown in the
previous papers of the author) are required detailed elaboration in a
separate paper.} \begin{eqnarray} \lefteqn{{\cal T}^{(1)}_{\mu\nu} =
\partial_\alpha \overline \psi_1 \gamma_{\alpha\mu} \partial_\nu \psi_1 +
\partial_\nu \overline \psi_1 \gamma_{\mu\alpha} \partial_\alpha \psi_1
+\nonumber}\\ &+&\partial_\alpha \overline \psi_2 \gamma_{\alpha\mu}
\partial_\nu \psi_2 +\partial_\nu \overline \psi_2 \gamma_{\mu\alpha}
\partial_\alpha \psi_2+{\cal L}^{(1)}\delta_{\mu\nu} \quad.
\end{eqnarray}
As a result the Hamiltonian is\footnote{The Hamiltonian can
also be obtained from the second-order Lagrangian presented
in~[13b,Eq.(18)] by means of the procedure developed by  M. V.
Ostrogradsky~\cite{OSTR} long ago (see also the Weinberg's remark on the
page B1325 of the first paper~\cite{WEIN}).  The Ostrogradsky's procedure
 seems not to have been applied in~\cite{AHLU1} to obtain conjugate
momentum operators.}
\begin{eqnarray}\label{H}
\lefteqn{{\cal H}^{(1)} =
\int \left [ -\partial_4 \overline \psi_2 \gamma_{4 4}\partial_4 \psi_2 +
\partial_i \overline \psi_2  \gamma_{ij} \partial_j \psi_2
-\right.\nonumber}\\ &-& \left.  \partial_4 \overline \psi_1 \gamma_{4 4}
\partial_4 \psi_1 + \partial_i \overline \psi_1 \gamma_{ij} \partial_j
\psi_1 +m^2 \overline \psi_1 \psi_1 -m^2 \overline \psi_2 \psi_2 \right ]
d^3 x \quad.  \end{eqnarray}
The quantized Hamiltonian \begin{equation}
{\cal H}^{(1)} =  \sum_\sigma \int \frac{d^3 p}{(2\pi)^3} E_p \, \left [
a_\sigma^{\dagger} (\vec p) a_\sigma (\vec p) +b_\sigma
(\vec p) b_\sigma^{\dagger} (\vec p) +c_\sigma^{\dagger} (\vec
p) c_\sigma(\vec p) +d_\sigma (\vec p) d^{\dagger}_\sigma
(\vec p)\right ]\quad, \end{equation}
is obtained after using the
plane-wave expansion following the procedure of, {\it e.g.},
refs.~\cite{ITZY,BOGO}.  Acknowledging the  suggestion of one critical
collegue I regard the matters of translational invariance and
positive-definiteness of the energy in the theory based on the ${\cal
L}^{(1)}$ in more detail. I proceed step by step to the fermionic
consideration of ref.~\cite[p.145]{ITZY}.\footnote{In order not to darken
the essence of the question I assume that  transitions $\psi_1
\leftrightarrow \psi_2$ and transitions between states of different
signs of energy (like in~\cite{ITZY}) are irrelevant at the moment.
Otherwise, the only correction should be taken into account where
necessary, namely, the commutators (\ref{c1},\ref{c2}) should be
generalized, see ref.~\cite{DVOE12}.} The condition of the translational
invariance imposes the constraints:
\begin{equation}
\psi_1 (x+a) = e^{-i P_\mu a_\mu}
\psi_1 (x) e^{i P_\mu a_\mu}\quad,\quad \psi_2 (x+a) = e^{-i P_\mu a_\mu}
\psi_2 (x) e^{i P_\mu a_\mu}\quad, \end{equation}
or in the differential form
\begin{eqnarray} \partial_\mu \psi_1 (x) &=& - i \left [ P_\mu , \psi_1
(x) \right ]_-\quad,\quad \partial_\mu \overline \psi_1 (x) = -i \left
[P_\mu , \overline \psi_1 (x) \right ]_- \quad,\quad\\ \partial_\mu \psi_2
(x) &=& - i \left [ P_\mu , \psi_2 (x) \right ]_-\quad,\quad \partial_\mu
\overline \psi_2 (x) = -i \left [ P_\mu , \overline \psi_2 (x) \right ]_-
\quad.\quad \end{eqnarray}
These constraints are satisfied provided that
\begin{eqnarray}
\left [ P_\mu , a_\sigma (\vec p) \right ]_- &=& - p_\mu a_\sigma (\vec p)
\quad,\quad
\left [ P_\mu , b_\sigma (\vec p) \right ]_- = - p_\mu b_\sigma (\vec
p)\quad,\quad\\
\left [ P_\mu , a_\sigma^\dagger (\vec p) \right ]_- &=& + p_\mu
a_\sigma^\dagger (\vec p) \quad,\quad
\left [ P_\mu , b_\sigma^\dagger (\vec p)
\right ]_- = + p_\mu b_\sigma^\dagger (\vec p)\quad.
\end{eqnarray}
Analogous relations exist for operators $c_\sigma (\vec p)$
and $d_\sigma (\vec p)$.
Replacing $P_\mu$ by its expansion, this is equivalent to
\begin{eqnarray}
a_\sigma^\dagger (\vec k) \left [ a_\sigma (\vec k), a_{\sigma^\prime}
(\vec p) \right ]_- + \left [ a_\sigma^\dagger (\vec k), a_{\sigma^\prime}
(\vec p) \right ]_- a_\sigma (\vec k) &=& -(2\pi)^3 \delta^{(3)} (\vec p
-\vec k) a_{\sigma^\prime} (\vec p)\\
b_\sigma (\vec k) \left [ b_\sigma^\dagger (\vec k), b_{\sigma^\prime}
(\vec p) \right ]_- + \left [ b_\sigma (\vec k), b_{\sigma^\prime}
(\vec p) \right ]_- b_\sigma^\dagger (\vec k) &=& -(2\pi)^3 \delta^{(3)}
(\vec p -\vec k) b_{\sigma^\prime} (\vec p)\\
a_\sigma^\dagger (\vec k) \left [ a_\sigma (\vec k),
a_{\sigma^\prime}^\dagger (\vec p) \right ]_- + \left [ a_\sigma^\dagger
(\vec k), a_{\sigma^\prime}^\dagger (\vec p) \right ]_- a_\sigma (\vec k)
&=& (2\pi)^3 \delta^{(3)} (\vec p -\vec k) a_{\sigma^\prime}^\dagger (\vec
p)\\
b_\sigma (\vec k) \left [ b_\sigma^\dagger (\vec k),
b_{\sigma^\prime}^\dagger (\vec p) \right ]_- + \left [ b_\sigma (\vec k),
b_{\sigma^\prime}^\dagger (\vec p) \right ]_- b_\sigma^\dagger (\vec k)
&=& (2\pi)^3 \delta^{(3)} (\vec p -\vec k) b_{\sigma^\prime} (\vec p)
\end{eqnarray}
We can list very similar formulas for the states
defined by the field function $\psi_2 (x)$. Therefore, we deduce the
commutation relations
\begin{eqnarray}\label{c1} \left [a_\sigma (\vec p),
a^\dagger_{\sigma^\prime} (\vec k) \right ]_{-} &=& \left [c_\sigma (\vec
p), c^\dagger_{\sigma^\prime} (\vec k) \right ]_{-} = (2\pi)^3
\delta_{\sigma\sigma^\prime}\delta (\vec p - \vec  k)\quad,\\
\label{c2}
\left [b_\sigma (\vec p), b^\dagger_{\sigma^\prime}
(\vec  k) \right ]_{-} &=&  \left [d_\sigma (\vec p),
d^\dagger_{\sigma^\prime} (\vec k) \right ]_{-} =
(2\pi)^3
\delta_{\sigma\sigma^\prime}\delta (\vec p - \vec  k)\quad.
\end{eqnarray}
It is easy to see that the Hamiltonian is  positive-definite and
the translational invariance still keeps  in the framework of this
description ({\it cf.} with ref.~\cite{AHLU1}). Please pay attention here:
{\it I did never apply the indefinite metric}, which is regarded to be a
rather obscure concept.

Analogously, from the definitions
\begin{eqnarray}
{\cal J}_{\mu} &=& -i \sum_i \left \{ \frac{\partial {\cal L}}{\partial
(\partial_\mu \phi_i)}
\phi_i -\overline \phi_i \frac{\partial {\cal L}}{\partial
(\partial_\mu \overline\phi_i)} \right \}\quad, \\
Q &=& -i \int {\cal J}_{4} (x)  d^3 x \quad,
\end{eqnarray}
and
\begin{eqnarray}\label{PL}
\lefteqn{{\cal M}_{\mu\nu,\lambda} = x_\mu {\cal T}_{\lambda\nu}
- x_\nu {\cal T}_{\lambda\mu} - \nonumber}\\
&-&i \sum_{i}
\left \{ \frac{\partial {\cal L}}
{\partial (\partial_\lambda \phi_i)} N_{\mu\nu}^{\phi_i} \phi_i +
\overline \phi_i  N_{\mu\nu}^{\overline\phi_i} \frac{\partial {\cal L}}
{\partial (\partial_\lambda \overline \phi_i)}\right \} \quad,\\
M_{\mu\nu} &=& -i \int  {\cal M}_{\mu\nu, 4} (x) d^3 x  \quad,
\end{eqnarray}
one can find the current operator
\begin{eqnarray}\label{qq1}
\lefteqn{{\cal J}^{(1)}_\mu = i \left [\partial_\alpha \overline \psi_1
\gamma_{\alpha\mu}
\psi_1 - \overline \psi_1 \gamma_{\mu\alpha} \partial_\alpha  \psi_1
+\right.\nonumber}\\
&+&\left.  \partial_\alpha\overline \psi_2 \gamma_{\alpha\mu}  \psi_2
- \overline \psi_2
\gamma_{\mu\alpha} \partial_\alpha \psi_2\right ]\quad,
\end{eqnarray}
and using (\ref{PL})  the spin momentum tensor
\begin{eqnarray}\label{ss1}
\lefteqn{S_{\mu\nu,\lambda}^{(1)} = i \left [ \partial_\alpha \overline \psi_1
\gamma_{\alpha\lambda} N_{\mu\nu}^{\psi_1} \psi_1 +  \overline \psi_1
N_{\mu\nu}^{\overline \psi_1}\gamma_{\lambda\alpha}
\partial_{\alpha} \psi_1 + \right.\nonumber}\\
&+& \left. \partial_\alpha \overline \psi_2
\gamma_{\alpha\lambda} N_{\mu\nu}^{\psi_2} \psi_2
+\overline \psi_2 N_{\mu\nu}^{\overline \psi_2}
\gamma_{\lambda\alpha} \partial_\alpha \psi_2 \right ]\quad.
\end{eqnarray}

If  the Lorentz group generators  (a $j=1$ case) are defined from
\begin{eqnarray}\label{lor}
&&\overline  \Lambda \gamma_{\mu\nu} \Lambda a_{\mu\alpha}
a_{\nu\beta} = \gamma_{\alpha\beta}\quad,\\
&& \overline \Lambda \Lambda =1\quad,\\ \label{lor3}
&&\overline \Lambda  = \gamma_{44} \Lambda^\dagger \gamma_{44}\quad.
\end{eqnarray}
then in order to keep the  Lorentz covariance of
the Weinberg
equations and of the Lagrangian (\ref{eq:Lagran1})
one should use the following generators:
\begin{equation}
N_{\mu\nu}^{\psi_1,  \psi_2 (j=1)} =
- N_{\mu\nu}^{\overline \psi_1 , \overline \psi_2 (j=1)} =
{1\over 6} \gamma_{5,\mu\nu}\quad,
\end{equation}
The matrix  $\gamma_{5,\mu\nu}= i \left [\gamma_{\mu\lambda},
\gamma_{\nu\lambda}\right ]_{-}$
is  defined to be Hermitian.
Let me note that  the matters of the choice of generators for Lorentz
transformations have also been regarded in~\cite{SANK}. Due to the fact
that the set of the Weinberg states is a degenerate set one can also
consider the situation when a Weinberg equation ({\it e.g.}, Eq.
(\ref{w1})) transfers over another one ({\it e.g.}, Eq. (\ref{w11})).
This case corresponds to the possibility of combining pure Lorentz
transformations with transformations of the inversion group; the
corresponding rules are different from (\ref{lor})-(\ref{lor3}) .

The quantized charge operator and the quantized  spin
operator follow immediately from (\ref{qq1}) and (\ref{ss1}):
\begin{equation}
Q^{(1)}=\sum_\sigma \int \frac{d^3 p}{(2\pi)^3} \, \left [
a_\sigma^\dagger (\vec p) a_\sigma (\vec p) -
b_\sigma (\vec p) b_\sigma^\dagger (\vec p) +
c_\sigma^\dagger (\vec p) c_\sigma (\vec p) -
d_\sigma (\vec p) d_\sigma^\dagger (\vec p)\right ]\quad,
\end{equation}
\begin{eqnarray}\label{spin}
\lefteqn{(W^{(1)}\cdot n) /m =   \sum_{\sigma\sigma^\prime}\int \frac{d^3
p}{(2\pi)^3}{1\over m^2 E_p} \overline u_1^\sigma (\vec p) (E_p \gamma_{44} -
i\gamma_{4i}p_i )\,\, I \otimes (\vec J \vec n)\,  u^{\sigma^\prime}_1
(\vec p) \times \nonumber}\\ & &\qquad \times \left [a_\sigma^\dagger (\vec p)
a_{\sigma^\prime} (\vec p) + c_\sigma^\dagger (\vec p) c_{\sigma^\prime} (\vec
p) - b_\sigma (\vec p) b_{\sigma^\prime}^\dagger (\vec p) -d_\sigma  (\vec p)
d_{\sigma^\prime}^\dagger  (\vec p) \right ]\quad
\end{eqnarray}
(provided that  the frame is chosen in such a way
that $\vec n \,\,\, \vert\vert \,\,\,\vec p$ \,\,\, is along the third axis).
It is easy to verify  the eigenvalues
of the charge operator are $\pm 1$,
and  of the Pauli-Lyuban'sky  spin operator are
\begin{equation}\label{heli}
\xi^*_\sigma (\vec J \vec n) \xi_{\sigma^\prime} = +1,\, 0\, -1
\end{equation}
in a massive case and $\pm 1$
in a massless case.\footnote{See the discussion of the massless limit of
the Weinberg bispinors in ref.~\cite{AHLU10,AHLU-PR}. While in a massless
limit $W_\mu n_\mu =0$ this does not signify that $W_\mu$ would be always
equal to zero; in this case we already cannot define a normalized
space-like vector $n_\mu$ whose space part is parallel to the vector
$\vec p$.  It becomes light-like.} Now we can answer the question:  why
``a queer reduction of degrees of freedom" did happen in the previous
papers~\cite{HAYA,KALB,AVDE}?  The origin of this surprising fact  follows
from the Hayashi (1973) paper, ref.~\cite[p.498]{HAYA}:  The requirement
of {\it ``that the physical realizable state satisfies  a quantal version
of the generalized Lorentz condition"}, formulas (18) of
ref.~\cite{HAYA},\footnote{Read:  ``a quantal version" of the  Maxwell
equations imposed on the state vectors in the Fock space.
Applying them leads to the case when Eq. (\ref{spin}) is equal to zero
{\it identically}.  Nonetheless, such a procedure should be taken
cautiously, see, {\it e.g.},  ref.~\cite[Table 2]{AHLU1}, for the
discussion of the acausal physical dispersion of the equations (4.19) and
(4.20) of ref.~[2b], ``which {\it are just Maxwell's free-space equations
for left- and right- circularly polarized radiation.}" See also the
footnote \# 1 in ref.~[28c].  Let me mention, the fact of existence of
`acausal' solutions is probably connected with the indefinite metric
problem, with the appearence of the ghost states in the gauge models and
with the concept of `action-at-a-distance', ref.~\cite{Chubyk}.} permits
one to eliminate upper (or down) part of the Weinberg ``bispinor" and to
remove transversal components of the remained part by means of the
``gauge" transformation (\ref{eq:gauge}), what {\it ``ensures the massless
skew-symmetric field is longitudinal".} The reader  can convince himself
in this ``obvious fact" by looking at the explicit form of the
Pauli-Lyuban'sky operator, Eq.  (\ref{spin}).  Taking into account both
positive- and negative- energy solutions ({\it cf.} with~\cite{AVDE}) in
the Lagrangian (\ref{eq:Lagran}) and {\it not} applying the generalized
Lorentz condition ({\it cf.} with~\cite{HAYA,KALB}) we are able  to
account for both transversal and longitudinal components, {\it i.e.}, to
describe a $j=1$ particle.  Furthermore, one can say even simpler: the
application of the generalized Lorentz condition may be successful to
the non-zero energy states of helicities $\pm 1$,\footnote{If the
energy is equal to zero,in my opinion, there is no any sense to speak
about helicity at all.} so in earlier works, as a matter of fact, the
authors implied the existence of such states.  On the other hand,
longitudinal components of the Weinberg fields are directly linked with
the mass of a $j=1$ particle, see~\cite{AHLU-PR} and, possibly, with the
concept of the ${\bf B}^{(3)}$ Evans-Vigier field~\cite{EVAN}.  This fact
can provide deeper understanding of relations between Casimir invariants
of a particle field and space-time structures. The presented wisdom does
{\it not} contradict with neither the Weinberg theorem nor the classical
limit, Eqs.  (\ref{cl1},\ref{cl2}) of the previous Section.  Thanks to the
mapping between the antisymmetric tensor and Weinberg formulations  the
conclusion is valid for both the Weinberg $2(2j+1)$ component ``bispinor"
and the antisymmetric (skew-symmetric) tensor field.  Thus, we have now
proven that a photon (a $j=1$ massless particle) can possess spin degrees
of freedom,  what is in accordance with experiment.  The contradictory
claims of several collegues about the pure ``longitudinal nature"  of
quantized antisymmetric fields, which they have been making since the
sixties and which are repeating until the present, are incredible and
unreasonable. We can suggest an analogy  considering the modified
electrodynamics recently proposed by Evans and Vigier. In fact, the
authors of the earlier ``longitudinal" papers ``align themselves" with the
concept of the ${\bf B}^{(3)}$ field (named it as the Kalb-Ramond field),
but, surprisingly, they reject transversal modes (after quantization)!? By
the way, it is obviously from the consideration of the similar construct
in the $(1/2,0)\oplus (0,1/2)$ representation that on an equal footing
those authors could claim that a $j=1/2$ massless neutrino field would be
pure longitudinal too\ldots Simply speaking, such claims are
absurdity\ldots

Finally, for the sake of completeness let me re-write Lagrangians
presented above  into the 12-component form:
\begin{equation}
{\cal L}^{(1)}  = - \partial_\mu \overline \Psi \Gamma_{\mu\nu}
\partial_\nu \Psi - m^2 \overline \Psi \Psi \quad, \end{equation} where
\begin{eqnarray}
\Psi = \pmatrix{\psi_1 \cr \psi_2}\quad,
\quad \overline \Psi = \pmatrix{\psi_1^\dagger & \psi_2^\dagger \cr}
\cdot \pmatrix{\gamma_{44}&0 \cr   0& -\gamma_{44}}
\end{eqnarray}
are the doublet wave functions,
\begin{eqnarray}
\Gamma_{\mu\nu} = \pmatrix{\gamma_{\mu\nu} &0 \cr 0& -\gamma_{\mu\nu}\cr}
\quad,\quad \Gamma^5 =\pmatrix{\openone &0\cr
0&-\openone\cr}\quad,\quad
\Gamma^0 =\pmatrix{0&\openone\cr
\openone &0\cr}\quad.
\end{eqnarray}
The Lagrangian ${\cal L}^{(2)}$ can be written in a similar fashion:
\begin{eqnarray}
\lefteqn{{\cal L}^{(2)} = - \partial_\mu^\dagger
\Psi^{(1)\,\dagger} \Gamma_{\mu\nu} \Gamma^5 \Gamma^0 \partial_\nu^\dagger
\Psi^{(2)}  -\partial_\mu \Psi^{(2)\,\dagger} \Gamma_{\mu\nu}
\Gamma^5 \Gamma^0 \partial_\nu \Psi^{(1)} -\nonumber}\\
&-& m^2 \Psi^{(1)\,\dagger} \Gamma^5 \Gamma^0 \Psi^{(2)}
+ m^2 \Psi^{(2)\,\dagger} \Gamma^5 \Gamma^0 \Psi^{(1)}\quad.
\end{eqnarray}

One can conclude this Section: the generalized Lorentz condition
can be incompatible with the specific properties of
the antisymmetric tensor field deduced from the ordinary approach of the
classical physics.  I mean that its application can lead (and did lead
in the  earlier papers) to  the loss of information about either
transversal or longitudinal modes of the antisymmetric tensor field. The
connection of the presented model with the Bargmann-Wightman-Wigner-type
quantum field theories deserves further elaboration. As a matter of fact
the presented model develops Weinberg and Ahluwalia ideas of the
Dirac-like description of bosons on an equal footing with fermions, {\it
i.e.}, on the ground of the $(j,0)\oplus (0,j)$ representation of the
Lorentz group.

\section{Weinberg propagators}

Accordingly  to the Feynman-Dyson-Stueckelberg ideas,
a causal propagator  has to be constructed
by using  the formula (e.~g., ref.~\cite[p.91]{ITZY})
\begin{eqnarray}
\lefteqn{S_F (x_2, x_1) =\int \frac{d^3 k}{(2\pi)^3}\frac{m}{E_k} \left [
\theta (t_2 -t_1) \, a \,\,
u^\sigma (k) \otimes \overline u^{\sigma} (k) e^{-ikx} + \right. }\nonumber\\
&&\left. \qquad\qquad  +\, \,\theta (t_1 - t_2) \, b \,\,
v^\sigma (k) \otimes \overline v^\sigma (k) e^{ikx} \right ] \quad ,
\end{eqnarray}
$x=x_2 -x_1$. In the $j=1/2$ Dirac theory  it results  to
\begin{equation}\label{dp}
S_F (x) = \int \frac{d^4 k}{(2\pi)^4} e^{-ikx} \frac{\hat k +m}{k^2 -m^2
+i\epsilon} \quad,
\end{equation}
provided that the constant $a$ and $b$ are determined by imposing
\begin{equation}
(i\hat \partial_2 -m) S_F (x_2, x_1) =\delta^{(4)} (x_2 -x_1) \quad,
\end{equation}
namely, $a=-b=1/ i$ .

However, in the framework of the Weinberg theory, ref.~\cite{WEIN},
which is a generalization of the Dirac ideas to higher spins,
the attempts of constructing a covariant propagator in such a way have been
fallen.  For example, on the page B1324 of ref.~[2a]\,   Weinberg writes:
``Unfortunately, the propagator arising from Wick's theorem is  {\it not}
equal to the covariant propagator except for $j=0$ and $j=1/2$. The
trouble is that the derivatives act on the $\epsilon (x) = \theta (x) -
\theta (-x)$ in $\Delta^C (x)$ as well as on the functions\footnote{In the
cited paper the following notation has been used:  $\Delta_1(x) \equiv i
\left [\Delta_+ (x) + \Delta_+ (-x)\right ]$\, ,\, $\Delta (x) \equiv
\Delta_+ (x) - \Delta_+ (-x)$ and $i\Delta_+ (x) \equiv \frac{1}{(2\pi)^3}
\int \frac{d^3 p}{2E_p} \exp (ipx)$.} $\Delta$ and $\Delta_1$. This gives
rise to extra terms proportional to equal-time $\delta$ functions and
their derivatives\ldots The cure is well known: \ldots compute the vertex
factors using only the original covariant part of [the Hamiltonian] ${\cal
H} (x)$; do not use [the Wick propagator] for internal lines; instead use
the covariant propagator, [the formula (5.8) in ref.~[2a]]." The
propagator, recently proposed in ref.~[35c,d] (see also other papers of
the same author), is the causal propagator.  ``Only the physically
acceptable causal solutions of the Weinberg equations enter these
propagators." However, it does not satisfy us down to the ground since the
old problem remains:  the Feynman-Dyson propagator is not the Green's
function of the Weinberg equation.  The covariant propagator presented
in~\cite{TUCK1}, while a Green's function of the $(1,0)\oplus (0,1)$
equation, would propagate kinematically spurious solutions~[35c]\ldots
The aim of the following work is to consider the problem of constructing
propagators in the framework of the model proposed in the previous
Sections.

The set of four equations has been proposed in Section II.
We consider the most general case.
Let us check, if the sum of four equations
($x=x_2 -x_1$)
\begin{eqnarray}\label{prop}
&&\hspace*{-1cm}\left [ \gamma_{\mu\nu} \partial_\mu \partial_\nu -m^2 \right ]
 \int  \frac{d^3 p}{(2\pi)^3 2E_p}
\left [ \theta (t_2 -t_1) \, a\,\,\,   {\cal U}_1^{\sigma\,(1)} (\vec p)
\otimes \overline {\cal U}_1^{\sigma\,(1)} (\vec p) e^{ipx}+\right
.\nonumber\\ &&\left.  \qquad\qquad+\theta (t_1 -t_2) \, b \,\,\, {\cal
V}_1^{\sigma\,(1)} (\vec p) \otimes \overline  {\cal V}_1^{\sigma\,(1)}
(\vec p) e^{-ipx} \right  ] +\nonumber\\ &+& \left [ \gamma_{\mu\nu}
\partial_\mu \partial_\nu + m^2 \right  ] \int \frac{d^3 p}{(2\pi)^3 2E_p}
\left [ \theta (t_2 -t_1) \, c\,\,\, {\cal U}_2^{\sigma\,(1)} (\vec p)
\otimes \overline {\cal U}_2^{\sigma\,(1)} (\vec p) e^{ipx}+ \right.
\nonumber\\ &&\left.  \qquad\qquad+\theta (t_1 -t_2) \, d \,\,\, {\cal
V}_2^{\sigma\,(1)} (\vec p) \otimes \overline  {\cal V}_2^{\sigma\,(1)}
(\vec p) e^{-ipx}\right  ] +\nonumber\\ &+&\left [ \widetilde
\gamma_{\mu\nu} \partial_\mu \partial_\nu + m^2 \right ] \int \frac{d^3
p}{(2\pi)^3 2E_p} \left [ \theta (t_2 -t_1) \, e\,\,\, {\cal
U}_1^{\sigma\,(2)} (\vec p) \otimes \overline {\cal U}_1^{\sigma\,(2)}
(\vec p) e^{ipx}+ \right.\nonumber\\ &&\left.  \qquad\qquad+\theta (t_1
-t_2) \, f \,\,\, {\cal V}_1^{\sigma\,(2)} (\vec p) \otimes \overline
{\cal V}_1^{\sigma\,(2)} (\vec p) e^{-ipx} \right ] +\nonumber\\ &+&\left
[\widetilde \gamma_{\mu\nu} \partial_\mu \partial_\nu - m^2 \right  ] \int
\frac{d^3 p}{(2\pi)^3 2E_p} \left [ \theta (t_2 -t_1) \, g\,\,\, {\cal
U}_2^{\sigma\,(2)} (\vec p) \otimes \overline {\cal U}_2^{\sigma\,(2)}
(\vec p) e^{ipx} +\right.\nonumber\\ &&\left.  \qquad\qquad+\theta (t_1
-t_2) \, h \,\,\, {\cal V}_2^{\sigma\,(2)} (\vec p) \otimes \overline
{\cal V}_2^{\sigma\,(2)} (\vec p) e^{-ipx} \right ] = \delta^{(4)} (x_2
-x_1)
\end{eqnarray}
can be satisfied by the definite choice of the
constant $a$, $b$ {\it etc}.  In the process of calculations  I  assume
that the set of the analogs of the ``Pauli spinors" in the $(1,0)$ or
$(0,1)$ spaces  is the complete set and it is normalized to
$\delta_{\sigma\sigma^\prime}$\, .

The simple calculations yield
\begin{eqnarray}
\lefteqn{\partial_\mu^{x_2} \partial_\nu^{x_2}  \left [ a\, \theta (t_2
-t_1)\, e^{ip(x_2 -x_1)} + b\, \theta (t_1 -t_2)\, e^{-ip(x_2 -x_1)}
\right ]=\nonumber}\\ &=& - \left [ a\, p_\mu p_\nu \theta (t_2 - t_1)
\exp \left [ ip(x_2 -x_1)\right ] + b\,  p_\mu p_\nu  \theta (t_1 -t_2)
\exp \left [ -ip (x_2 -x_1) \right ] \right ] + \nonumber\\ &+& a\left [ -
\delta_{\mu 4} \delta_{\nu 4} \delta^{\,\,\prime} (t_2 -t_1) +i (p_\mu
\delta_{\nu 4} +p_\nu \delta_{\mu 4}) \delta (t_2 -t_1) \right ] \exp
\left [i \vec p (\vec x_2 - \vec x_1)\right ] +\nonumber\\ &+& b\, \left [
\delta_{\mu 4} \delta_{\nu 4} \delta^{\,\,\prime} (t_2 -t_1) + i (p_\mu
\delta_{\nu 4} +p_\nu \delta_{\mu 4}) \delta (t_2 -t_1) \right ] \exp
\left [-i\vec p (\vec x_2 - \vec x_1)\right ] \quad;
\end{eqnarray}
and
\begin{eqnarray}
{\cal U}_1^{(1)}\overline {\cal U}_1^{(1)} ={1\over 2} \pmatrix{m^2
\openone & S_p \otimes S_p\cr \overline S_p \otimes \overline S_p
&m^2 \openone\cr}\quad,\quad {\cal U}_2^{(1)}\overline {\cal U}_2^{(1)} =
{1\over 2}\pmatrix{-m^2 \openone & S_p \otimes S_p\cr \overline S_p
\otimes \overline S_p &-m^2 \openone\cr}\quad, \end{eqnarray}
\begin{eqnarray} {\cal U}_1^{(2)}\overline {\cal U}_1^{(2)} ={1\over 2}
\pmatrix{-m^2 \openone & \overline S_p \otimes \overline S_p\cr S_p
\otimes S_p &-m^2\openone\cr}\quad,\quad {\cal U}_2^{(2)}\overline {\cal
 U}_2^{(2)} = {1\over 2} \pmatrix{m^2 \openone & \overline S_p \otimes
 \overline S_p\cr S_p \otimes  S_p &m^2\openone\cr}\quad, \end{eqnarray}
where
\begin{eqnarray} S_p &=& m + (\vec J \vec p) +\frac{(\vec J \vec
 p)^2}{E+m}\quad,\\ \overline S_p &=& m - (\vec J \vec p) + \frac{(\vec J
\vec p)^2}{E+m}\quad.  \end{eqnarray}
Due to the fact that
\begin{eqnarray}
\left [E- (\vec J \vec
p)\right ]  S_p \otimes S_p &=& m^2 \left [ E+ (\vec J \vec p)\right
]\quad,\quad\\
\left [E+ (\vec J \vec p)\right ] \overline S_p \otimes
\overline S_p &=& m^2 \left [ E- (\vec J \vec p)\right ]\quad
\end{eqnarray}
after
simplifying the left side of (\ref{prop}) and comparing it with the right
side we find: the  causal  propagator is admitted by using the ``Wick's
formula" for the time-ordered particle operators provided that the
constants are equal to $1/ 4im^2$.  It is necessary to  consider all four
equations, Eqs.  (\ref{w1},\ref{w2},\ref{w11},\ref{w21}).

The $j=1$ analogs of the formula (\ref{dp})  for the Weinberg propagators
follows
from the formula (3.6) of ref.~[35d] immediately:
\begin{equation}\label{propa1}
S_F^{(1)} ( p ) = -\frac{1}{i(2\pi)^4  (p^2  +m^2 -i\epsilon)} \left [
\gamma_{\mu\nu} p_\mu p_\nu   -  m^2  \right ]\quad,
\end{equation}
\begin{equation}\label{propa2}
S_F^{(2)} ( p ) = -\frac{1}{i(2\pi)^4  (p^2  +m^2 -i\epsilon)} \left [
\gamma_{\mu\nu} p_\mu p_\nu   +  m^2  \right ]\quad,
\end{equation}
\begin{equation}\label{propa3}
S_F^{(3)} ( p ) = -\frac{1}{i(2\pi)^4  (p^2  +m^2 -i\epsilon)} \left [
\widetilde\gamma_{\mu\nu} p_\mu p_\nu   +  m^2  \right ] \quad,
\end{equation}
\begin{equation}\label{propa4}
S_F^{(4)} ( p ) = -\frac{1}{i(2\pi)^4  (p^2  +m^2 -i\epsilon)} \left [
\widetilde \gamma_{\mu\nu} p_\mu p_\nu   -  m^2  \right ] \quad.
\end{equation}

The conclusions are: one can construct an analog of the Feynman-Dyson
propagator for the $2(2j+1)$ model and, hence, a ``local"
theory provided that the Weinberg states are
``quadrupled"  in the $j=1$ case. They cannot propagate separately each
other ({\it cf.} with the Dirac $j=1/2$ case).

\section{Massless limit: Can the 6-component
Weinberg-Tucker-Hammer equations describe the electromagnetic field?}

In previous Sections
the equivalence of the Weinberg-Tucker-Hammer approach and the Proca
approach for describing $j=1$ states has been found.
The  $2(2j+1)$ component wave functions are given
by Eq. (\ref{eq:EH}) and by the formulas obtained after applying
inversion group operations to (\ref{eq:EH}).  The aim of the present
Section is to consider the question, under which conditions the
Weinberg-Tucker-Hammer $j=1$ equations can be transformed to Eqs.  (4.21)
and (4.22) of ref.~[2b]~:
\begin{eqnarray} {\bf \nabla}\times \left [
{\vec E} -i {\vec B}\right ] +i (\partial/\partial t) \left [{\vec E}
-i{\vec B} \right ] &=&0\quad, \qquad (4.21) \nonumber\\ {\bf
\nabla}\times \left [ {\vec E} +i {\vec B}\right ] -i (\partial/\partial
t) \left [{\vec E} +i{\vec B} \right ] &=&0\quad.  \qquad (4.22) \nonumber
\end{eqnarray} By using the bivector interpretation of $\psi$ (in the
chiral representation) and the explicit forms of the
Barut-Muzinich-Williams matrices, We are able to recast the $j=1$
Tucker-Hammer equation  (\ref{eq:Tucker}) which is free of tachyonic
solutions, or the Proca equation, Eq. (\ref{eq:eq}) of the Section II, to
the form \begin{eqnarray}\label{ME0} m^2 E_i &=& - {\partial^2 E_i \over
\partial t^2} +\epsilon_{ijk} {\partial \over \partial x_j} {\partial B_k
\over \partial t} + {\partial \over \partial x_i} {\partial E_j \over
\partial x_j}\quad,\\ \label{ME1} m^2 B_i &=& \epsilon_{ijk} {\partial
\over \partial x_j} {\partial E_k \over \partial t} +  {\partial^2 B_i
\over \partial x_j^2} -{\partial \over \partial x_i} {\partial B_j \over
\partial x_j}\quad.  \end{eqnarray} The Klein-Gordon equation (the
D'Alembert equation in the massless limit) \begin{equation}\label{DA}
\left (\frac{\partial^2}{\partial t^2} - \frac{\partial^2}{\partial
x_i^2}\right ) F_{\mu\nu} = - m^2 F_{\mu\nu} \end{equation} is implied
($c=\hbar=1$). Introducing  vector operators we write equations in the
following form:
\begin{eqnarray}\label{MY1} {\partial \over \partial t}\,
\mbox{curl}\, {\vec B} + \mbox{grad}\, \mbox{div}\, {\vec E} - {\partial^2
{\vec E} \over \partial t^2} &=& m^2 {\vec E}\quad,\\
\label{MY2} {\bf
\nabla}^2 {\vec B} -\mbox{grad}\, \mbox{div}\, {\vec B} + {\partial \over
\partial t}\, \mbox{curl}\, {\vec E} &=& m^2 {\vec B}\quad.
\end{eqnarray}
Taking into account the definitions:  \begin{eqnarray}
\rho_e &=&\mbox{div} \, {\vec E}\quad, \quad {\vec J}_e = \mbox{curl} \,
{\vec B} - {\partial {\vec E} \over \partial t} \quad, \label{ME11}\\
\rho_m &=& \mbox{div} \,{\vec B} \quad, \quad {\vec J_m} = - {\partial
{\vec B} \over \partial t} - \mbox{curl}\, {\vec E} \quad,\label{ME12}
\end{eqnarray} relations of the vector algebra (${\vec X}$ is an
arbitrary vector):  \begin{equation} \mbox{curl}\, \mbox{curl}\, {\vec X}
= \mbox{grad}\, \mbox{div}\, {\vec X} -\nabla^2 {\vec X}\quad,
\end{equation}
and the Klein-Gordon equation (\ref{DA}) we obtain two
equivalent sets of equations, which complete the Maxwell's set.  The first
one is \begin{eqnarray} && {\partial {\vec J}_e \over \partial t}
+\mbox{grad} \,\rho_e = m^2  {\vec E}\quad,\label{mm1}\\ && {\partial
{\vec J}_m \over \partial t} +\mbox{grad} \, \rho_m =0\quad;\label{mm2}
\end{eqnarray} and the second one is \begin{eqnarray} &&\mbox{curl}\,
{\vec J}_m =0\label{mm3}\\ &&\mbox{curl} \,{\vec J}_e = -m^2 {\vec
B}\quad.\label{mm4} \end{eqnarray}
One can  obtain the equations in
different unit systems after one recalls, {\it e.g.}, relations of the
Appendix of ref.~\cite{JACKSON}.  I would also like to remind that the
Weinberg set of equations (and, hence, the equations
(\ref{mm1}-\ref{mm4})\footnote{Beginning with the dual massive equations
and setting $\vec C \equiv \vec E$, \, $\vec D \equiv \vec B$ we could
obtain \begin{eqnarray} &&{\partial {\vec J}_e \over \partial t}
+\mbox{grad} \,\rho_e = 0\quad,\\ && {\partial {\vec J}_m
\over \partial t} + \mbox{grad} \,\rho_m =m^2 \vec B\quad;
\end{eqnarray}
and
\begin{eqnarray} && \mbox{curl}\, \vec J_e =0\quad,\\ &&\mbox{curl}\,
\vec J_m = m^2 \vec E\quad.  \end{eqnarray}
This would signify that  the
physical content spanned by massive dual fields would be different. The
reader can easily reveal parity-conjugated equations from Eqs.
(\ref{w11},\ref{w21}).}) can be obtained on the basis of a very few number
of postulates; in fact, by using the Lorentz transformation rules for the
Weinberg bivector (or for the antisymmetric tensor field) and the
Ryder-Burgard relation~\cite{AHLU1,AHLU2,DVOE2,DVOE3,DVOE3A}.

In a massless case the situation is different. Firstly,
the set of equations (\ref{ME12}), with the left side are chosen to
be zero, is ``an identity satisfied by certain space-time derivatives
of $F_{\mu\nu}$\ldots, namely, refs.~\cite{DYSO,TANI,HORW1}.
\begin{equation}
{\partial F_{\mu\nu} \over \partial x^\sigma} +
{\partial F_{\nu\sigma} \over \partial x^\mu} +
{\partial F_{\sigma\mu} \over \partial x^\nu} = 0\quad."
\end{equation}
I believe that the similar consideration for the dual field $\widetilde
F_{\mu\nu}$ as in refs.~\cite{DYSO,TANI} can reveal that the same is
true for the first equations (\ref{ME11}). So, in the massless case we
come across the problem of interpretation of the charge and currents.

Secondly, in order to  satisfy the massless equations (\ref{mm3},\ref{mm4})
one should assume that the currents are represented in the gradient forms
of some scalar fields $\chi_{e,m}$.  What physical significance have these
chi-functions?  In the massless case the charge densities are  (see
equations (\ref{mm1},\ref{mm2})) \begin{equation} \rho_e = -
\frac{\partial \chi_e}{\partial t} + const \quad, \quad \rho_m = -
\frac{\partial\chi_m}{\partial t} + const \quad,
\end{equation}
what tells  us that $\rho_e$ and $\rho_m$ are constants provided that the
primary functions $\chi_{e,m}$ are linear functions in time (decreasing or
increasing?).  It is useful to compare the resulting equations for
$\rho_{e,m}$ and ${\vec J}_{e,m}$ and the fact of appearence of the
functions $\chi_{e,m}$ with the 5-potential formulation of electromagnetic
theory~\cite{TANI}, see also
refs.~\cite{BOYA,HORW1,GERS,BAND,MALI,HORW2}.  I believe, this concept
can also be useful for explanation of the $E=0$ solutions in higher-spin
equations~\cite{OPPEN,GIAN,AHLU10} which have been ``baptized" by
Moshinsky and Del Sol in~\cite{MOSH} as `relativistic cockroach nest'.
Next, I would like to note the following. We can obtain
the Maxwell's free-space equations,
in the definite choice  of the  $\chi_e$ and $\chi_m$, namely,
in the case they are constants. In ref.~\cite{GERS}
it was mentioned that solutions of Eqs. (4.21,4.22) of ref.~[2b]
satisfy the equations of the type (\ref{ME0},\ref{ME1}), {\it ``but not
always vice versa".} Interpretation of this statement
and investigations of Eq. (\ref{eq:Tucker})  with different initial and
boundary conditions (or of the functions $\chi$) deserve further
elaboration (both theoretical and experimental).

The question also arises on the
transformation of the field function (\ref{eq:EH}) from one to another
frame.  I would like to draw your attention at the remarkable fact which
follows from a consideration of the problem in the momentum
representation.  For the first sight, one could conclude that under a
transfer from one to another frame one has to describe the field by  the
Lorentz transformed function $\psi^\prime ({\bf p}) = \Lambda ({\bf
p})\psi ({\bf p})$.  However, if take into account the possibility of
combining the Lorentz, dual (chiral) and parity transformations in the
case of higher spin equations\footnote{This possibility has been
discovered earlier and investigated in~\cite{AHLU1}.}  and that all the
equations for the four functions (\ref{w1}), (\ref{w2}), (\ref{w11}) and
(\ref{w21}) reduce to the equations for ${\bf E}$ and ${\bf B}$, which
appear to be the same in a massless limit, one could come to a different
situation.  The four bispinors ${\cal U}_1^{\sigma\, (1)} ({\bf p})$,
${\cal U}_2^{\sigma\, (1)} ({\bf p})$, ${\cal U}_1^{\sigma\, (2)} ({\bf
p})$ and ${\cal U}_2^{\sigma\, (2)} ({\bf p})$, see Eqs.  (\ref{bb1}),
(\ref{bb2}), (\ref{b11}) and (\ref{b21}), form a complete set (as well as
the transformed ones $\Lambda ({\bf p}) {\cal U}_i^{\sigma\, (k)} ({\bf
p}))$ for each value of $\sigma$.

Namely,
\begin{eqnarray} \label{comp}
\lefteqn{a_1 {\cal U}_1^{\sigma\, (1)} ({\bf p})
\overline {\cal U}_1^{\sigma\, (1)} ({\bf p}) + a_2 {\cal U}_2^{\sigma\,
(1)} ({\bf p}) \overline {\cal U}_2^{\sigma\, (1)} ({\bf p}) +\nonumber}\\
&+& a_3 {\cal U}_1^{\sigma\, (2)} ({\bf p}) \overline {\cal U}_1^{\sigma\,
(2)} ({\bf p}) + a_4 {\cal U}_2^{\sigma\, (2)} ({\bf p}) \overline
{\cal U}_2^{\sigma\, (2)} ({\bf p}) = \openone \quad.
\end{eqnarray}
Constants $a_i$ are defined by the choice of the normalization of
bispinors.  In any other frame we are able to obtain the primary wave
function by choosing appropriate coefficients $c_i^k$ of the expansion of
the wave function (in fact, using appropriate dual rotations and
inversions)
\begin{equation} \Psi ({\vec p})= \sum_{i,k=1,2} c_i^k {\cal
U}_i^{(k)} ({\vec p})\quad.
\end{equation}
The same statement should be
valid for negative-energy solutions, since their explicit forms coincide
with the ones of positive-energy bispinors in the case of the
Hammer-Tucker formulation for a $j=1$ boson, ref.~\cite{TUCK1}.  Using the
plane-wave expansion one can prove this conclusion in the coordinate
representation.  Thus, the question of what we observe in the experiment
would be solved depending on the fixing of the relative phase factor
between left- and right- parts of the field function (between $\vec E$
and $\vec B$, indeed) by appropriate physical conditions  we are
interested.

At last, I have to note that the massless case reveals a very strange
thing.\footnote{I am grateful to Dr. A. E. Chubykalo for pointing out
this fact and for discussions.} The massless equations
(\ref{mm3},\ref{mm4}) written in the integral form lead to a conclusion
about $\oint \vec J_{e,m} \cdot d\vec l =0$. This is obviously unacceptable
from a viewpoint of experiment. Thus, we have to conclude that either
the $j=1$ field cannot be massless  or  there exist hidden parameters
which all field functions (and, probably, space-time characteristics)
depend on.

Finally, let me mention that in the nonrelativistic limit $c\rightarrow
\infty$ one obtains the dual
Levi-Leblond's ``Galilean Electrodynamics",
refs.~\cite{LEVY,CRAW}.

The main conclusion of the paper is:\footnote{This conclusion also follows
from the results of the
paper~\cite{AHLU1,AHLU10,DVOE10,DVOE11,DVOE12,DVOE13,DVOE14,EVAN,DVOE96}
and ref.~[2b] provided that the fact that $({\vec J} {\vec p})$ has no the
inverse matrix  has been taken into account.} The Weinberg-Tucker-Hammer
massless equations (or the Proca equations for $F_{\mu\nu}$), see also
(\ref{ME0}) and (\ref{ME1}), are equivalent to the Maxwell's equations in
the definite choice of the initial and boundary conditions, what proves
their consistency.  Their massless limit were shown in ref.~\cite{AHLU10}
to be free of kinematical acausalities as opposed to Eqs. (4.21) and
(4.22) of ref.~[2b]. The Weinberg-Tucker-Hammer approach permits us to
clarify the question of the claimed `longitudinal nature' of the
antisymmetric tensor field. It is free of the problem of the indefinite
metric in the Fock space. The $j=1$ bosons are considered in a very
similar fashion as fermions in the Dirac approach.  This provides a
convenient mathematical formalism for discussing properties of the $j=1$
bosons with respect to discrete symmetries operations. Therefore, we have
to agree with S.  Weinberg who spoke out about the equations (4.21) and
(4.22):  {\it ``The fact that these field equations are of first order for
any spin seems to me to be of no great significance\ldots"}~[2b,p.  B888].
In the meantime, I would not like to darken  theories based on the use of
the vector potentials, {\it i.e.}, of the $D(1/2,1/2)$ representation of
the Lorentz group. While the description of the $j=1$ massless field
using this representation contradicts with the Weinberg theorem
$B-A=\lambda$, what signifies that we do not have well-defined creation
and annihilation operators in the beginning of a quantization procedure,
one cannot forget about significant achievements of these theories. The
formalism proposed here could be helpful only if we shall necessitate to
go beyond the framework of the Standard Model, {\it i.e.} if we shall come
across the reliable experimental results which cannot have satisfactory
explanation on the basis of the concept of a minimal coupling introduced
in the conventional manner~(see, {\it e.g.}, ref.~\cite{AHLU2} for a
discussion of the neutrino model which forbids such a form of the
interaction).

Many questions related with the problem of longitudinal modes of the $j=1$
field, their relations with tachyonic models  (particularly, with the
concept of the Action At a Distance and the Recami's Extended Relativity),
with the problem of the interpretations of mass and spin, with the
problem of gauge degrees of freedom as well remain in the field of our
future researches.

\section{Acknowledgments}

I am glad to express my gratitude to Prof. D. V. Ahluwalia for
useful advice which permitted to begin the work in this direction.
His answers on my questions were very incentive to
the fastest writing the paper. The firm support that my thoughts are not
groundless was in one of the letters, he sent to me.  The papers of
Professor  D. V. Ahluwalia in respectable journals led me to the ideas
presented here and in my previous publications.

A considerable part of this paper has been
written in August, 1994.  The distribution of the ideas, presented above,
in many simposiums and in private correspondence has provided with
additional confirmations that the Maxwell's electromagnetic theory should
be looked by `fresh glance' and that some problems of both classical and
quantum theory are urgently required adequate explanation. Recently I
learnt about the papers of other authors published in the recent years in
many journals over the world.  The  papers of H.  A.  M\'unera and O.
Guzm\'an on the longitudinal solutions of relativistic wave equations, of
A. E. Chubykalo and R.  Smirnov-Rueda~\cite{Chubyk} on the
`action-at-a-distance' concept in the classical electrodynamics ({\it cf.}
with the researches of F. J.  Belinfante in QED, refs.~\cite{BELIN}), the
papers of E.  Recami, W.  Rodrigues and J.  Vaz on superluminal phenomena,
which have recently been observed, also deserve steady attention and
verifications as well.    Furthermore, as I also learnt recently, some of
issues considered  by me in the present essay have been touched in the
old works~\cite{BELIN0,OHAN}.  My papers present themselves some
insights in the mentioned problems, which are though connected with the
previous considerations but, in my opinion, more relevant to the modern
field theory.   The group-theoretical basis for our
researches has been proposed in the old papers of E.  Wigner, Y. S.  Kim
and S.  Weinberg.

Now I know that many physicists put forward
``inconvenient" questions earlier.  As for me, I heard for the first time
about limitations of the Maxwell's electromagnetic theory (with all
consequences for other well-known models) in 1984 from Professor A. F.
Pashkov.  I want to thank him too.

We acknowledge the efforts of the editors of some respectable journals.
Their support ensures that these researches will
continue.  My papers of the relevant nature have been published in:  Nuovo
Cim.  A108 (1995) 1467; ibid 111B (1996) 483; Int.  J. Theor. Phys. 34
(1995) 2467; ibid 35 (1996) 115; Sov. J. Nucl. Phys. 48 (1988) 1770; Rev.
Mex. Fis. Suppl.  40 (1994) 352; ibid 41 (1995) 159; Hadronic J. 16 (1993)
423; ibid 16 (1993) 459; Hadronic J.  Suppl.  10 (1995) 349; ibid 10
(1995) 359; Russ. Phys. J. 37 (1994) 898; Bol.  Soc.  Mex. Fis. 8-3 (1994)
113; ibid 9-3 (1995) 28; Proc.  ICTE'95, Cuautitl\'an, Sept.  1995; Proc.
IV Wigner Symp., Guadalajara, Aug.  1995; Inv.  Cient\'{\i}fica (1996), in
press.  Several papers are now under the review process in various
journals.

I greatly appreciate valuable discussions with Profs. R. N.
Faustov, I.  G.  Kaplan, Y.~S.~ Kim, A.  Mondragon, M.~Moreno, M.
Moshinsky, E.  Recami, S.  Roy, M.  Sachs, N.  B.  Skachkov,
Yu.~F.~Smirnov, J. Vaz and G. Ziino.

I am grateful to Zacatecas University,  M\'exico, for a Full Professor
position. This work has been partly supported by el Mexican Sistema
Nacional de Investigadores, el Programa de Apoyo a la Carrera Docente
and by the CONACyT, M\'exico under the research project 0270P-E.\\

\end{document}